\documentclass[preprint,
preprintnumbers,amsmath,amssymb,nofootinbib]{revtex4}
\usepackage{etex}
\usepackage{amssymb,amsthm,amscd,amsbsy,array}
\usepackage{bm}
\usepackage{soul} 
\usepackage{graphics,graphicx,xcolor}
\usepackage{soul} 
\usepackage{algorithm}
\usepackage{algorithmic}
\usepackage{longtable}
\usepackage{graphicx}
\usepackage{epstopdf}
\usepackage{subfigure}
\usepackage{float}
\usepackage[utf8]{inputenc}
\usepackage[T1]{fontenc}
\usepackage[english]{babel}
\usepackage{graphicx}
\usepackage{float}
\usepackage{tikz}
\usepackage{xcolor}
\usepackage[utf8]{inputenc}
\usepackage[T1]{fontenc}
\usepackage[english]{babel}
\usepackage{graphicx}
\usepackage{float}
\usepackage{tikz}
\usepackage{xcolor}
\usepackage{makecell}

\usepackage[colorlinks=true, pdfstartview=FitV, linkcolor=blue, citecolor=blue, urlcolor=blue]{hyperref} 
\newcommand{\cyan}{\textcolor{cyan}}
\newcommand{\magenta}{\textcolor{magenta}}
\newcommand{\red}{\textcolor{red}}

\newcommand{\orange}{\textcolor{orange}}

\newcommand{\gb}{\quad\colorbox{green}}

\newcommand{\dgreen}{\textcolor[rgb]{0,0.5,0}}

\newenvironment{redtext}{\color{red}}
{\ignorespacesafterend}
\newenvironment{bluetext}{\color{blue}}{\ignorespacesafterend}

\newenvironment{magentatext}{\color{magenta}}{\ignorespacesafterend}
\newenvironment{cyantext}{\color{cyan}}{\ignorespacesafterend}
\newenvironment{orangetext}{\color{orange}}
{\ignorespacesafterend}

\newcommand{\bmagenta}{\begin{magentatext}}
\newcommand{\emagenta}{\end{magentatext}}
\newcommand{\bcyan}{\begin{cyantext}}
\newcommand{\ecyan}{\end{cyantext}}
\newcommand{\bblue}{\begin{bluetext}}
\newcommand{\eblue}{\end{bluetext}}
\newcommand{\bred}{\begin{redtext}}
\newcommand{\ered}{\end{redtext}}
\newcommand{\borange}{\begin{orangetext}}
\newcommand{\eorange}{\end{orangetext}}

\numberwithin{equation}{section}

\let\ssection=\section
\renewcommand{\section}{\setcounter{equation}{0}\ssection}
\newcommand{\beq}{\begin{equation}}
\newcommand{\eeq}{\end{equation}}

\def\aand{{\quad\text{\small and}\quad}}
\def\where{{\quad\text{\small where}\quad}}

\def\s.t.{{\quad\text{\small such that}\quad}}

\newcommand{\oor}{\quad\text{\small or}\quad}

\newcommand{\bX}{{\mathbf{X}}}

\newcommand{\grad}{{\vnabla}}

\newcommand{\SO}{\mathrm{SO}}

\def\vnabla{{\bm{\nabla}}}

\def\smallover\#1/\#2{\hbox{$\textstyle\frac{\#1}{\#2}$}} %

\def\Ort{{\rm O}}
\def\orth{{\rm o}}

\def\benu{\begin{enumerate}}
\def\eenu{\end{enumerate}}
\def\bitem{\begin{itemize}}
\def\eitem{\end{itemize}}

\def\beq{\begin{equation}}
\def\eeq{\end{equation}}
\def\beqa{\begin{eqnarray}}
\def\eeqa{\end{eqnarray}}
\def\nn{\nonumber}
\def\barray{\left(\begin{array}}
\def\earray{\end{array}\right)}
\def\barraynb{\begin{array}}
\def\earraynb{\end{array}}
\def\ort{{\rm o}}
\def\Ort{{\rm O}}

\def\?{{\quad\gb{\fbox{\texttt{?}}\;}}\quad}

\def\p{{\partial}}

\def\v0{\mathbf{0}}

\def\p{\partial}

\def \p{{\partial}}

\newcommand{\const}{\mathop{\rm const.}\nolimits}

\def\smallover#1/#2{\hbox{$\textstyle\frac{#1}{#2}$}} %
\def\smallcirc{{\raise 0.5pt \hbox{$\scriptstyle\circ$}}}
\def\cabove(#1){\stackrel{\smallcirc}{#1}}
\def\ccabove(#1){\stackrel{\smallcirc\smallcirc}{#1}}
\def\cccabove(#1){\stackrel{\,\smallcirc\smallcirc\smallcirc}{#1}\,}
\def\2{{\smallover1/2}}

\newcommand{\bigbox}[1]{\fbox{%
\rule[-20pt]{0pt}{45pt}$\;\;\displaystyle{#1}\;\;$}
}

\let\ssection=\section
\renewcommand{\section}
{\setcounter{equation}{0}\ssection}

\def\besub{\begin{subequations}}
\def\esub{\end{subequations}}

\begin{document}
\preprint{arXiv:2403.02230 
} 

\title{Conformally related vacuum gravitational waves and their symmetries
}

\author{
Q. L. Zhao$^{1}$\footnote{mailto: zhaoqliang@mail2.sysu.edu.cn},
P. M. Zhang$^{1}$\footnote{mailto: zhangpm5@mail.sysu.edu.cn},
and
P. A. Horv\'athy$^{2}$\footnote{mailto: horvathy@lmpt.univ-tours.fr}
}

\affiliation{
${}^1$ School of Physics and Astronomy, Sun Yat-sen University, Zhuhai, (China)
\\
${}^2$  Institut Denis-Poisson CNRS/UMR 7013 - Universit\'e de Tours - Universit\'e d'Orl\'eans Parc de Grammont, 37200; Tours, France \\
}
\date{\today}

\begin{abstract}
A special conformal transformation which carries a vacuum gravitational wave into another vacuum one is built by using M\"obius-redefined time. It can  either  transform a globally defined vacuum wave into a vacuum sandwich  wave, or carry the gravitational wave into itself.  The first type, illustrated by linearly and circularly polarized vacuum plane gravitational waves, permutes the symmetries and the geodesics. Our second type is a pp wave with conformal $\Ort(2,1)$ symmetry, which seem to have escaped attention so far, is an anisotropic generalization of the familiar inverse-square profile. An example inspired by molecular physics, for which the particle can escape, or perform periodic motion, or fall into the singularity is studied in detail.
\end{abstract}

\maketitle

\tableofcontents

\section{Introduction}

We consider special class of gravitational wave (GW) space-times whose metric is written in Brinkmann coordinates \cite{Brinkmann} as, 
\begin{eqnarray}
ds^2=dX^2+dY^2+2dUdV-2H(U,\mathbf{X})\,dU^2\,.
\label{pp-wave}
\end{eqnarray}
Here
$U$ and $V$ are light-cone coordinates, $X$ and $Y$ represent the transverse plane and $H(U,\mathbf{X})$ is the wave profile.
Brinkmann space-times are smooth Lorentz manifolds endowed  with a covariantly constant null Killing vector field $\xi=\partial_V$ \cite{DGH91}.
Such structures arised before in the study of the one-parameter central extension of the Galilei group \cite{Barg54} called  the \emph{Bargmann group} \cite{DBKP}. In the proposed  Kaluza-Klein-type  ``Bargmann''  framework  \cite{DBKP,DGH91,Eisenhart,Burdet1985}  the motions in ordinary space are obtained by projecting out the ``vertical'' null direction along the coordinate $V$ and  identifying the other null coordinate, $U$, with Newtonian time.  The profile $H(U,\mathbf{X})$ is the Newtonian potential \cite{Eisenhart,Burdet1985,DGH91}. 
\goodbreak

An insight is provided by the so-called {memory effect} \cite{ZelPol,Braginsky1985,Grishchuk1989,Ehlers, Sou73} 
which amounts to studying test particles initially at rest  by using the symmetries of the corresponding background space-time. 
It is particularly convenient to use conformal symmetries \cite{Carroll4GW,Zhang:2017rno,Zhang:2017geq,Zhang:2018srn,Marsot:2021tvq,EZHrev} 
generated by  conformal Killing vectors (CKV),
\begin{eqnarray}
\mathcal{L}_Wg_{ab}=2\psi g_{ab}\,,
\label{L-D}
\end{eqnarray}
where $\psi$ is a smooth function of the coordinates \cite{Bondi:1958aj,Sippel1986,exactsol,MaMa,Keane:2004dpc}. 
In   flat Minkowski space-time, the conformal Lie algebra of CKVs is $15$ dimensional. 
The same number of dimensions arises for conformally flat space-times, which include, in addition to Minkowski space-time, also that for oscillator and for a linear potential  \cite{Niederer1973,Burdet1985,Silagadze2021,ZZH2022,DGH91,Penrose2022,Dunajski23,Sen:2022vig}~. 

The maximal number of symmetries of a non-conformally-flat space-time is  $7$ \cite{Keane:2004dpc,MaMa,exactsol}. Their research is simplified when the space-time is conformally related to one whose symmetries are known, and therefore the CKVs are interchanged. This happens, in particular, for the time redefinition \eqref{Gibbons-trans} below, proposed in \cite{Keane:2004dpc,Gibbons2014}.  
 
In this paper we study special time-redefined conformal transformations of  simple rational form\ ref{Mobius} referred to as  M\"obius transformations. They can  (i) either interchange \emph{two vacuum GWs} (as illustrated by linearly polarized plane GWs (LPP) and circularly polarized vacuum plane GWs (CPP) \cite{Zhang2022,Elbistan:2022umq,Masterov}, or (ii) leave the wave form-invariant (Secs. \ref{secIII} and \ref{secIV}) as illustrated by a wave inspired by the anisotropic   polar molecule
 \cite{Camblong:2001zt,Camblong:2003mz,Moroz:2009nm}.

The $U$-dependence brought in by  M\"obius transformation  is ``mild'', though, because of the simple rational form of \eqref{Mobius}.
 Realistic GWs are however  ``spikes" : they are ``sandwich waves'' 
  \cite{Hawking} with a short wave zone
$[U_i,U_j]$ outside of which the space-times is flat \cite{Hawking,BoPi89,Zhang:2017rno,Zhang:2017geq,Steinbauer1997,Podolsky:1998in,PodSB,Zhang:2017jma,SBComment}. 

Approximate sandwich waves with rapidly decaying amplitude can be considered by putting an $U$-dependent factor into the profile. In this paper we limit our attention to simple Gaussians, although more general forms can (and have been) also considered \cite{Hawking,Zhang:2017geq}.  
Their properties are studied in sec. \ref{sandwichSec}.

Some entertaining historical facts are recounted in sec.\ref{historysec}.

\section{Conformal transformations 
of gravitational waves} \label{secII}

The GW space-time, $ds^2$ in \eqref{pp-wave}, can be transformed into another GW, $d\widetilde{s}^2$, by  a special conformal transformation with redefined time  $f(\widetilde{U})$ \cite{Gibbons2014,Keane:2004dpc}\,,
\begin{eqnarray}
U=f(\tilde{U}),
\qquad
\mathbf{X}=\sqrt{\cabove(f)}\,\mathbf{\widetilde{X}},
\qquad
V=\tilde{V}-\frac{1}{4}\frac{\ccabove(f)}{\cabove(f)}\,\mathbf{\widetilde{X}}^2\,,
\label{Gibbons-trans}
\end{eqnarray}
where $(\cabove(\cdot))$ means $d/d\widetilde{U}$.
The corresponding conformal relation is,
\begin{eqnarray}
&&ds^2=\Omega^2d\widetilde{s}^2=\cabove(f)d\widetilde{s}^2\,, \label{conf-relation} 
\\[4pt]
&&d\widetilde{s}^2=d\widetilde{\mathbf{X}}^2+2d\widetilde{U}d\widetilde{V}-2\widetilde{H}(\widetilde{U},\widetilde{\mathbf{X}})d\widetilde{U}^2 \,, \label{c-new-GW} 
\\[4pt]
&&\widetilde{H}(\widetilde{U},\widetilde{\mathbf{X}})=\cabove(f)\!H\Big[\widetilde{\mathbf{X}}\sqrt{\cabove(f)},\, f(\widetilde{U})\Big]+\frac{1}{4}\mathcal{S}_{\widetilde{U}}(f)\,\widetilde{\mathbf{X}}^2,
\label{confprof}
\end{eqnarray}
where $\mathcal{S}_{\widetilde{U}}(f)$ is the Schwarzian derivative,
\beq
\mathcal{S}_{\widetilde{U}}(f)
=\frac{\cccabove(t)}{\cabove(t)}
-\frac{3}{2}\left(\frac{\ccabove(t)}{\cabove(t)}\right)^2\,.
\label{Sch}
\eeq

The vacuum condition for a pp-wave space-time \eqref{pp-wave} requires the Ricci tensor to vanish, $R_{\mu\nu}=0$, which implies that
\begin{eqnarray}
\Delta{H}=H_{,XX}+H_{,YY}=0\,. 
\label{vacuum condition}
\end{eqnarray}
This condition involves only the spatial behavior of the wave profile.  
Assuming that $H$ is that of a vacuum,  $\widetilde{H}$ in \eqref{confprof} will satisfy also the vacuum condition if the Schwarzian derivative term vanishes \cite{Masterov,ZZH2022}, which yield in turn  
 a \emph{special M\"obius conformal transformation} (SMCT),
\besub
\begin{align}
U&=f(\widetilde{U})=\frac{A\widetilde{U}+B}{C\widetilde{U}+D}\,,
 \label{ABCD-U}
\\[6pt]
 V&=\widetilde{V}+\frac{1}{2}\frac{C}{C\widetilde{U}+D}\widetilde{\mathbf{X}}^2\,, 
 \quad \mathbf{X}=\Omega\,\widetilde{\mathbf{X}}\,,
 \;\where\;
 \Omega=\frac{\sqrt{AD-BC}}{C\widetilde{U}+D}\,.
\label{M-trans}
\end{align}
\label{Mobius}
\esub
$A$, $B$, $C$ and $D$ here are arbitrary constants such that $AD-BC\neq0$. 
Under such a redefinition  the metric \eqref{c-new-GW} becomes,
\begin{eqnarray}
ds^2\,=\,\Omega^2\,
d\widetilde{s}^2
\,=\, d\widetilde{\mathbf{X}}^2+2d\widetilde{U}d\widetilde{V}
-2\Omega^2
{H}\left(f(\widetilde{U})
\,,\Omega\,\widetilde{\mathbf{X}}
\right)
d\widetilde{U}^2\,.
\label{s-c-new-GW}
\end{eqnarray}

The new wave profile  is in general different from the initial one.
An initially $U$-independent profile (as e.g. for Brdicka   \eqref{usual-LPP}) becomes indeed $U$-dependent. 
Examples will be seen in section \ref{secIII}.

However it might happen also that the  wave is \emph{invariant} under 
M\"obius redefinition --- i.e., \eqref{Mobius} acts  as a \emph{symmetry}
 studied in some detail in sec. \ref{secIV} and 
 illustrated by the pp wave inspired by molecular physics and studied in sect.\ref{PolarSec}.

\section{Conformally related vacuum gravitational waves}\label{secIII}

Hence we  focus our attention at vacuum plane GWs with line element
\begin{eqnarray}
ds^2=dX^2+dY^2+2dUdV-\Big[\alpha(U)(X^2-Y^2)+2\gamma(U)XY\Big]\,dU^2,
\label{PP-GW}
\end{eqnarray}
where the arbitrary functions
$\alpha(U)$ and $\gamma(U)$ correspond to the  ``$+$'' and ``$\times$'' polarization modes. These waves are taken conformally into an approximate sandwich form \eqref{s-c-new-GW}  by \eqref{ABCD-U}-\eqref{M-trans} \cite{Zhang:2017rno,Zhang:2017geq,Podolsky:1998in,Zhang:2017jma,Andrzejewski:2018pwq} 
with new profile function
\beq
-2\widetilde{H}=\Omega^4\left[\widetilde{\alpha}(\widetilde{U})(\widetilde{X}^2-\widetilde{Y}^2)+2\widetilde{\gamma}(\widetilde{U})\widetilde{X}\widetilde{Y}\right]. 
\label{confH}
\eeq
where $\widetilde{\alpha}(\widetilde{U})=\alpha[f(\widetilde{U})]$ and
$\widetilde{\gamma}(\widetilde{U})=\gamma[f(\widetilde{U})]$.
The new GWs include two classes which correspond to different choices of the coefficients $A$, $B$, $C$ and $D$.
$C=0$ means  a dilation and an $U$-translation of the original GWs which does not bring any new insight and will therefore not considered further.

\begin{figure}[htb]
\centering
\subfigure[\label{Delta1}]{
\begin{minipage}[t]{0.54\linewidth}
\centering
\includegraphics[width=6.5cm]{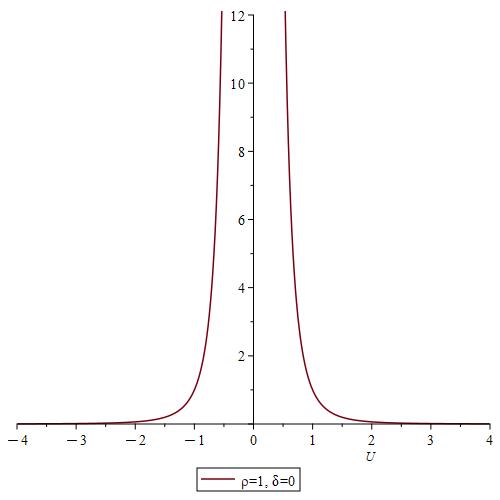}
\end{minipage}
}
\subfigure[\label{Delta2}]{
\begin{minipage}[t]{0.42\linewidth}
\centering
\includegraphics[width=6.5cm]{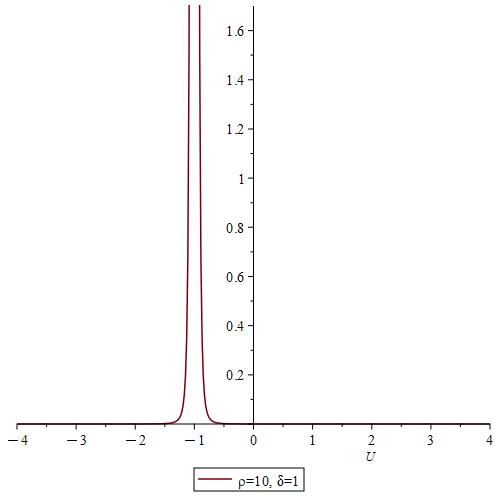}
\end{minipage}
}
\caption{\textit{\small The conformal factor \eqref{N(U)} determines the width and position of the wave \eqref{S-Gibbons}: \ref{Delta1} is for  parameters $\rho=1$ and $\delta=0$  and FIG. \ref{Delta2} 
is
 for $\rho=10$ and $\delta=1$,  respectively.}
\label{Deltas}}
\end{figure}

$C\neq0$ introduces in turn a new, rationally-redefined  scale factor. In terms of the redefined parameters $\rho=C/\sqrt{AD-BC}$ and $\delta=D/C$ which determine the width and the center of the new GW shown in FIG. \ref{Deltas}, the conformal factor
can  be presented as
\beq
\Omega^4=
\frac{1}{\left[\rho(U+\delta)\right]^4}\,.
\label{Nfactor}
\eeq
Apart of focusing and shifting, the parameters $\rho$ and $\delta$ do not change the trajectory. Choosing
  $\rho=1$ and $\delta=0$ for the sake of simplicity,
\beq
\Omega^4(U)
= \frac{1}{U^4}\,
\label{N(U)}
\eeq
generates the special rational transformation  \cite{Keane:2004dpc},
\begin{eqnarray}
U=-\frac{1}{\widetilde{U}}\,, \quad \mathbf{X}=\frac{\widetilde{\mathbf{X}}}{\widetilde{U}}\,,
\quad V=\widetilde{V}+\frac{\widetilde{\mathbf{X}}^2}{2\widetilde{U}}\,.
\label{S-Gibbons}
\end{eqnarray}

Eqns  \# (2.10) and \#  (2.11) of Andrzejewski and al.
\cite{Andrzejewski:2018pwq,Andrzejewski:2018zby} are also
similar, except for that their  profiles tend, unlike ours, to a Dirac delta. 

The M\"obius mapping SMCT \eqref{ABCD-U}-\eqref{M-trans} shown in FIG.\ref{Deltas}  {shrinks a globally defined  
 GW  into one concentrated around a single point} which then  behaves as an approximate sandwich wave \cite{Zhang:2017rno,Zhang:2017geq,Podolsky:1998in,Zhang:2017jma,Andrzejewski:2018pwq}. 
 
 Eqn. \eqref{S-Gibbons} is in fact the oscillator counterpart of the conformal transformation applied to planetary motion with a time-dependent gravitational constant, proposed by Dirac \cite{Dirac38,DGH91}. 

Keane and Tupper \cite{Keane:2004dpc} noted in particular that \eqref{S-Gibbons} allows us to obtain a  conformally related ``dual'' space-time. Our M\"obius-redefined time  and SMCT \eqref{ABCD-U}-\eqref{M-trans} have this property also, since the inverse transformation is identical to the original one.

The general vacuum GWs in \eqref{PP-GW} admit Killing vectors of the form
\begin{eqnarray}
	\widehat{\beta}=\bm{\beta}\,\partial_{\bm{X}}-\dot{\beta}_iX^i\partial_V, 
\label{CCKV}
\end{eqnarray}
where the two-vector $\bm{\beta}=(\beta_i)$ satisfies the vectorial Sturm-Liouville equation \cite{Torre,SL-C}:
\begin{eqnarray}
	\ddot{\beta}_i(U)=K_{ij}\beta_j(U), \quad \mathrm{\small with} \quad  K_{ij}=\left(
	\begin{array}{cc}
		\alpha(U) & \gamma(U) \\
		\gamma(U) & -\alpha(U)
	\end{array}\right)\,, 
\label{SturmLiouville}
\end{eqnarray}
where $\dot{(\cdot)}$ means $d/dU$.
The transformation \eqref{Mobius} then  
 carries the Killing vector \eqref{CCKV} into :
\begin{eqnarray}
	\widehat{\widetilde{\beta}}=\widetilde{\bm{g}}(\widetilde{U})\p_{\widetilde{\mathbf{X}}}-\widetilde{\mathbf{X}}\cdot\cabove(\widetilde{\bm{g}})\p_{\widetilde{V}}\,, 
\where \;
	\widetilde{\bm{g}}(\widetilde{U})=\Omega^{-1}\bm{g}\left(f(\widetilde{U})\right)\,.
	\label{M-CCKV}
\end{eqnarray}
$\widetilde{\bm{g}}$ here satisfies the redefined Sturm-Liouville equation, 
\begin{eqnarray}
	\ccabove(\widetilde{{g}})_j+\widetilde{K}_{ij}\widetilde{{g}}^i, 
	\quad \mathrm{with} \quad \widetilde{K}_{ij}=\Omega^4\left(
	\begin{array}{cc}
		\widetilde{\alpha}(\widetilde{U}) & \widetilde{\gamma}(\widetilde{U}) \\
		\widetilde{\gamma}(\widetilde{U}) & -\widetilde{\alpha}(\widetilde{U})
	\end{array}\right).
\end{eqnarray}

Below we illustrate our point by two vacuum GWs, one linearly polarized, and the other circularly polarized. Both are globally defined and have a 7-dimensional symmetry algebra. Then we study how their symmetries and geodesics change under the M\"obius transformation \eqref{S-Gibbons}.

\subsection{Conformally related linearly polarized vacuum GWs}\label{ConfRelLP}

\benu
\item
The simplest globally defined \emph{linearly polarized vacuum} GW 
(LPP) of Brdicka  \cite{Brdicka}, whose metric is,
\begin{eqnarray}
d{s}^2=d{X}^2+d{Y}^2+2d{U}d{V}-({X}^2-{Y}^2)\,d{U}^2\,. 
\label{usual-LPP}
\end{eqnarray}
Its CKVs are obtained by solving the conformal Killing equations,
 \begin{eqnarray}
    	W_{L}=\eta\partial_U+\left(2\rho V+\epsilon-\bm{X}\cdot\frac{d\bm{g}_L}{dU}\right)\partial_V 
    	+(\rho\bm{X}+\bm{g}_{L})\cdot\partial_{\bm{X}}\,, 
\label{CKV-LPP}
    \end{eqnarray}
where 
\begin{eqnarray}
\mathbf{g}_{L}(U)&=&
(\delta_1\sin U+\beta_1\cos U)\mathbf{e}_{X}+(\delta_2\cosh U-\beta_2\sinh U)\mathbf{e}_{Y}\,.
\label{Carroll-V-L-1}
\end{eqnarray}
Here  $\eta$, $\epsilon$, $\rho$, $\delta_i$ and $\beta_i$ are arbitrary constants which generate time-translations, $\widehat{E}$, vertical-translations, $\widehat{N}$, dilations, $\widehat{D}$, space-translations, $\widehat{P}_i$, and boosts $\widehat{G}_i$, respectively.  These symmetries span the 7-dimensional homothetic algebra $\mathcal{E}_7$,
\begin{eqnarray}
&&[\widehat{P}_i,\widehat{P}_j]=0, \quad [\widehat{G}_i,\widehat{G}_j]=0, \quad [\widehat{P}_i,\widehat{G}_j]=\delta_{ij}\bar{\omega}\widehat{N}, \quad [\widehat{D},\widehat{N}]=-2\widehat{N}, \nonumber  \\
&&[\widehat{D},\widehat{G}_i]=-\widehat{G}_i, \quad [\widehat{D},\widehat{P}_j]=-\widehat{P}_j, \quad [\widehat{D},\widehat{E}]=0, \quad [\widehat{E},\widehat{G}_i]=-\widehat{P}_i, \nonumber \\
&&[\widehat{E},\widehat{P}_1]=\widehat{G}_1, \quad [\widehat{E},\widehat{P}_2]=-\widehat{G}_2\,. 
\label{LPP-Lie-1}
\end{eqnarray}
The 
$\bm{g}_L$-terms in \eqref{CKV-LPP} can be collected into
\begin{eqnarray}
    	\bm{g}_{L}\cdot\partial_{\bm{X}}-\dot{\bm{g}_L}\bm{X}
	\cdot\partial_V\,,
	\label{III.9}
\end{eqnarray}
which is \eqref{CCKV}. 

\item
The conformal transformation \eqref{S-Gibbons} carries the Brdicka wave \eqref{usual-LPP} into a \emph{rational  LPP} 
 with  damped profile,
\begin{eqnarray}
d\widetilde{s}^2=d\widetilde{X}^2+d\widetilde{Y}^2+2d\widetilde{U}d\widetilde{V}-\frac{1}{\widetilde{U}^4}(\widetilde{X}^2-\widetilde{Y}^2)d\widetilde{U}^2. 
\label{LPP-pulse}
\end{eqnarray}
whose CKVs can be obtained either directly or  by the conformal transformation \eqref{S-Gibbons},
\begin{eqnarray}
	&&\widetilde{W}_{L}=\eta\widetilde{U}^2\partial_{\widetilde{U}}+\left(2\rho \widetilde{V}+\epsilon-\eta\frac{1}{2}\widetilde{\bm{X}}^2
	-\widetilde{\bm{X}}\cdot\cabove(\widetilde{\bm{g}})_{L}\right)\partial_{\widetilde{V}} 
	+\left(\rho \widetilde{\bm{X}}+\eta\widetilde{U}\widetilde{\bm{X}}+\widetilde{\bm{g}}_{L}\right)\cdot\partial_{\widetilde{\bm{X}}}\,, 
	\qquad\qquad
	\label{LPP-pulse-CKV}
\end{eqnarray}
where
\begin{eqnarray}
&&\widetilde{\mathbf{g}}_{L}(\widetilde{U})= 
\widetilde{U}\left(-\delta_1\sin \frac{1}{\widetilde{U}}+\beta_1\cos\frac{1}{\widetilde{U}}\right)\mathbf{e}_{X}
+\widetilde{U}\left(\delta_2\cosh\frac{1}{\widetilde{U}}+\beta_2\sinh\frac{1}{\widetilde{U}}\right)\mathbf{e}_{Y}\,.
\qquad\qquad
 \label{Carroll-V-L-2}
\end{eqnarray}

Note for further reference that the $\bm{\widetilde{g}}_L$-terms in \eqref{LPP-pulse-CKV}  combine into a solution of \eqref{M-CCKV}.

The parameters represent the same symmetries as in the Brdicka case except for $\eta$, which becomes a special Killing vector (SCKV) identified as an expansion $\widehat{K}$, 
\begin{eqnarray}
\widehat{K}
=U^2\p_U-\frac{1}{2}\mathbf{X}^2\p_V+U\mathbf{X}\cdot\p_{\mathbf{X}}\,,
\label{Khat}
\end{eqnarray}
which acts as a redefined-time translation $\widehat{E}=\p_{U}$.
The commutation relations are,
\begin{eqnarray}
&&[\widehat{P}_i,\widehat{P}_j]=0, \quad [\widehat{G}_i,\widehat{G}_j]=0, \quad [\widehat{P}_i,\widehat{G}_j]=\delta_{ij}\bar{\omega}\widehat{N}, \quad [\widehat{D},\widehat{N}]=-2\widehat{N}\,, 
\nonumber  \\
&&[\widehat{D},\widehat{G}_i]=-\widehat{G}_i, \quad [\widehat{D},\widehat{P}_j]=-\widehat{P}_j, \quad [\widehat{D},\widehat{K}]=0, \quad [\widehat{K},\widehat{G}_i]=-\widehat{P}_i\,, 
\nonumber \\
&&[\widehat{K},\widehat{P}_1]=\widehat{G}_1, \quad [\widehat{K},\widehat{P}_2]=-\widehat{G}_2\,. \label{LPP-Lie-2}
\end{eqnarray}
Thus the algebra $\mathcal{E}_7\supset\mathcal{G}_6$ for the Brdicka GW \eqref{usual-LPP} is transformed, for the rational-time LPP GW \eqref{LPP-pulse},
into
\beq
\mathcal{S}_7\supset\mathcal{E}_6\supset\mathcal{G}_5\,.
\eeq
Here $\mathcal{S}$, $\mathcal{E}$, $\mathcal{G}$ are the  special conformal algebra,  homothetic algebra and isometric algebra generators, respectively. The subscripts indicate the dimension of the algebra. The commutation relations do not change even if the CKVs do \cite{Elbistan:2020ffe,Keane:2004dpc}.

\item The \emph{circularly polarized (CPP) GW} has line element
\begin{eqnarray}
ds^2=dX^2+dY^2+2dUdV-\Big[\cos(2\omega U)(X^2-Y^2)+2\sin(2\omega U)XY\Big]dU^2, \qquad\quad
\label{CPP-metric}
\end{eqnarray}
where $\omega$ is an arbitrary constant frequency. The corresponding CKVs were studied, e.g., in \cite{Zhang:2018srn,Elbistan:2022umq}~: 
\begin{eqnarray}
W_{C}=\eta\left[\partial_U+\omega\left(X\p_{Y}-Y\p_{X}\right)\right]+\left(2\rho V+\epsilon-\bm{X}\cdot\dot{\bm{g}}_C\right)\partial_V 
+(\rho\bm{X}+\bm{g}_{C})\cdot\partial_{\bm{X}}\,,\qquad 
\end{eqnarray}
where
\begin{eqnarray}
&&\mathbf{g}_{C}(U)=g_{C1}(U)\,\mathbf{e}_{X}+g_{C2}(U)\,\mathbf{e}_{Y}, \label{Geo-CPP-1}  \\ [8pt]
&&g_{C1}(U)=\beta_2\mathcal{D}_U\left(\sin{\omega}U\cdot\sin\omega_-U\right)+\delta_2\mathcal{D}_U\left(\sin{\omega}U\cdot\cos\omega_-U\right) \nonumber  \\ [4pt]
&&\qquad +\beta_1\mathcal{D}_U\left(\cos{\omega}U\cdot\sin\omega_+U\right)-\delta_1\mathcal{D}_U\left(\cos{\omega}U\cdot\cos\omega_+U\right), \\ [8pt]
&&g_{C2}(U)=-\beta_2\mathcal{D}_U\left(\cos{\omega}U\cdot\sin\omega_-U\right)-\delta_2\mathcal{D}_U\left(\cos{\omega}U\cdot\cos\omega_-U\right) \nonumber \\ [4pt]
&&\qquad +\delta_1\mathcal{D}_U\left(\sin{\omega}U\cdot\cos\omega_+U\right)-\beta_1\mathcal{D}_U\left(\sin{\omega}U\cdot\sin\omega_+U\right)\,,
\end{eqnarray}
where $\omega_{\pm}=\sqrt{\omega^2\pm1}$ and $\mathcal{D}_U$ is the bilinear derivative $\mathcal{D}_U(f\cdot g)=g\frac{df}{dU}-f\frac{dg}{dU}$\,. These formulae represent also analytic geodesics in the CPP GW space-time, as said before.

The parameters $\eta$, $\rho$, $\epsilon$, $\delta_i$ and $\beta_i$ generate  ``screw'' symmetries  $\widehat{S}$ \cite{Elbistan:2022umq}, namely dilations $\widehat{D}$, vertical-translations $\widehat{N}$, space-translations $\widehat{P}_i$, and boosts $\widehat{G}_i$, respectively, span a 7-d  homothetic algebra $\mathcal{E}_7\supset\mathcal{G}_6$ \cite{Zhang:2018srn}.

\item Inserting  \eqref{S-Gibbons} into \eqref{CPP-metric}
yields the \emph{rational CPP GW} whose line element is,
\begin{eqnarray}
d\widetilde{s}^2&=&d\widetilde{X}^2+d\widetilde{Y}^2+2d\widetilde{U}d\widetilde{V}
\nn \\[8pt]
&&-\frac{1}{\widetilde{U}^4}
\left[\cos\left(\frac{2\omega}{\widetilde{U}}\right)(\widetilde{X}^2-\widetilde{Y}^2)+2\sin\left(\frac{2\omega}{\widetilde{U}}\right)\widetilde{X}\widetilde{Y}\right]d\widetilde{U}^2.\qquad
 \label{CPP-pulse-metric}
\end{eqnarray}
 Its CKVs are obtained as in the rational-time LPP case, 
 \begin{eqnarray}
&&\widetilde{W}_{C}=\eta\left[\partial_{\widetilde{U}}+\widetilde{U}\widetilde{\bm{X}}\cdot\p_{\widetilde{\bm{X}}}+\omega\left(\widetilde{X}\p_{\widetilde{Y}}-\widetilde{Y}\p_{\widetilde{X}}\right)\right] 
\nonumber \\[6pt]
&&\qquad\; 
+\left(2\rho \widetilde{V}+\epsilon-\bm{\widetilde{X}}\cdot\cabove(\widetilde{\bm{g}})_{C}\right)\partial_{\widetilde{V}}+(\rho\bm{\widetilde{X}}+\bm{\widetilde{g}}_{C})\cdot\partial_{\bm{\widetilde{X}}}\,, 
\label{CKV-rCPP}
\end{eqnarray}
where
\begin{eqnarray}
&&\widetilde{\mathbf{g}}_{C}(\widetilde{U})=\widetilde{g}_{C1}(\widetilde{U})\mathbf{e}_{\widetilde{X}}+\widetilde{g}_{C2}(\widetilde{U})\mathbf{e}_{\widetilde{Y}}, \label{Geo-CPP-2} \\ [8pt]
&&\widetilde{g}_{C1}=
-\beta_2\frac{\widetilde{U}}{\omega_+}\mathcal{D}_{\widetilde{U}}\left(\widetilde{U}\cos\frac{\omega_+}{\widetilde{U}}\cdot\widetilde{U}\cos\frac{\omega}{\widetilde{U}}\right)
-\delta_2\frac{\widetilde{U}}{\omega_+}\mathcal{D}_{\widetilde{U}}\left(\widetilde{U}\sin\frac{\omega_+}{\widetilde{U}}\cdot\widetilde{U}\cos\frac{\omega}{\widetilde{U}}\right) \nonumber  \\ [4pt]
&&\qquad
\,+\beta_1\frac{\widetilde{U}}{\omega}\mathcal{D}_{\widetilde{U}}\left(\widetilde{U}\cos\frac{\omega_-}{\widetilde{U}}\cdot\widetilde{U}\sin\frac{\omega}{\widetilde{U}}\right)
-\delta_1\frac{\widetilde{U}}{\omega}\mathcal{D}_{\widetilde{U}}\left(\widetilde{U}\sin\frac{\omega_-}{\widetilde{U}}\cdot\widetilde{U}\sin\frac{\omega}{\widetilde{U}}\right), \\ [8pt]
&&\widetilde{g}_{C2}=
-\beta_2\frac{\widetilde{U}}{\omega_+}\mathcal{D}_{\widetilde{U}}\left(\widetilde{U}\cos\frac{\omega_+}{\widetilde{U}}\cdot\widetilde{U}\sin\frac{\omega}{\widetilde{U}}\right)
-\delta_2\frac{\widetilde{U}}{\omega_+}\mathcal{D}_{\widetilde{U}}\left(\widetilde{U}\sin\frac{\omega_+}{\widetilde{U}}\cdot\widetilde{U}\sin\frac{\omega}{\widetilde{U}}\right) \nonumber  \\ [4pt]
&&\qquad
+\beta_1\frac{\widetilde{U}}{\omega}\mathcal{D}_{\widetilde{U}}\left(\widetilde{U}\cos\frac{\omega_-}{\widetilde{U}}\cdot\widetilde{U}\cos\frac{\omega}{\widetilde{U}}\right)
-\delta_1\frac{\widetilde{U}}{\omega}\mathcal{D}_{\widetilde{U}}\left(\widetilde{U}\sin\frac{\omega_-}{\widetilde{U}}\cdot\widetilde{U}\cos\frac{\omega}{\widetilde{U}}\right).
\end{eqnarray}
are also analytical geodesics in the rational CPP GW space-time.

\begin{figure}[htb]
  \centering
  \subfigure[\label{sk1}]{
  \begin{minipage}[t]{0.4\linewidth}
  \centering
  \includegraphics[width=6.2cm]{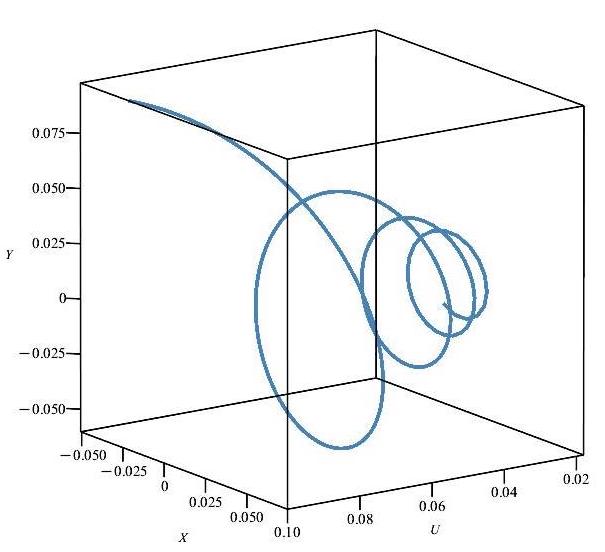}
  \end{minipage}
  }
  \subfigure[\label{sk2}]{
  \begin{minipage}[t]{0.4\linewidth}
  \centering
  \includegraphics[width=5.4cm]{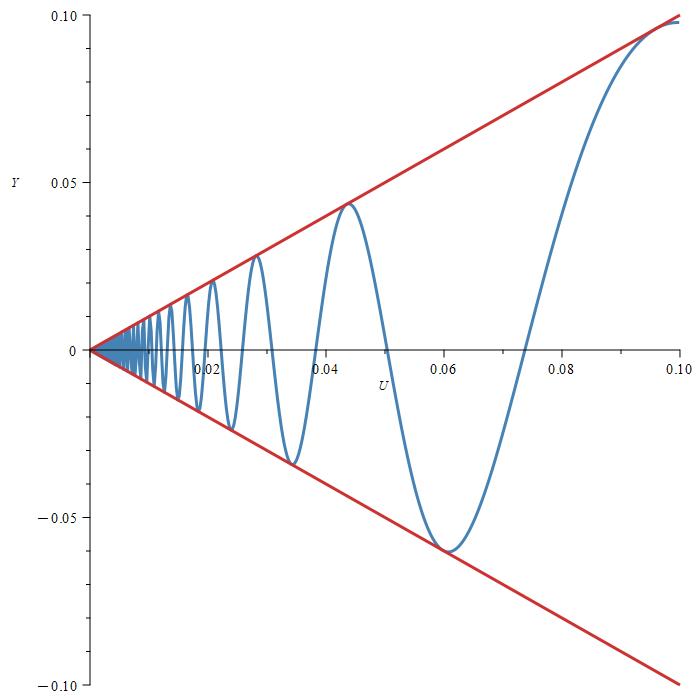}
 \end{minipage}
  }
 \\
 \caption{\textit{\small
\ref{sk1}: the ``screw'' \eqref{growingscrew} of the rational CCP GW \eqref{CPP-pulse-metric} expands linearly with $U$.
   FIG. \ref{sk2} shows its projection onto the $Y-U$ plane.}
 \label{sk}
  }
\end{figure}

Here the parameters $\rho$, $\epsilon$, $\delta_i$, $\beta_i$ represent the same symmetries as for the CPP wave, --- except for $\eta$, which is a new special symmetry denoted by $\widehat{S}_{K}$,
\begin{eqnarray}
\widehat{S}_{K}=\widetilde{U}^2\p_{\widetilde{U}}-\frac{1}{2}\widetilde{\mathbf{X}}^2\p_{\widetilde{V}}
+\widetilde{U}\widetilde{\mathbf{X}}\cdot\p_{\widetilde{\mathbf{X}}}+\omega(\widetilde{X}\p_{\widetilde{Y}}-\widetilde{Y}\p_{\widetilde{X}})\,, \label{Se}
\end{eqnarray}
which corresponds to Eq. \# (147) of Keane and Tupper in Ref. \cite{Keane:2004dpc}. Its geometric meaning is obtained by integrating the Killing vector \eqref{Se}. 
Its space part,
\beq\left\{
\barraynb{lll}
X&=&-\frac{U}{U_0}\left[X_0\cos\left(\frac{\omega(U-U_0)}{UU_0}\right)
+Y_0\sin\left(\frac{\omega(U-U_0)}{UU_0}\right)\right] 
\\[12pt]
Y&=&\;\;\frac{U}{U_0}\left[Y_0\cos\left(\frac{\omega(U-U_0)}{UU_0}\right)
-X_0\sin\left(\frac{\omega(U-U_0)}{UU_0}\right)\right]\,
\earraynb\right.\,,
\label{growingscrew}
\eeq
where $X_0$, $Y_0$ and $U_0$ are initial positions, 
 describes a ``growing screw'' whose size increases
linearly with $U$ while its frequency  decreases as shown in FIG. \ref{sk}. 
A similar ``screw'' has also been found for planetary motions with for time-dependent gravitational constant in Newtonian gravity  \cite{Dirac38,DGH91}.
\eenu

\subsection{Gedesics, found numerically and analytically}\label{analsols}

Analytic solutions are readily derived from the results  in section \ref{secIII}. 
Our clue is that the {Sturm-Liouville} equation \eqref{SturmLiouville} 
 for \emph{symmetries} is indeed \emph{identical} to the \emph{equations of motion} satisfied by the transverse coordinates $\bX(U)$ \cite{Lukash},
\beq
\ddot{X}_i(U)=K_{ij}X_j(U), 
\label{SLbis}
\eeq
(while the  3rd component $V(U)$ is then obtained by horizontal lift \cite{DBKP,DGH91}).

\begin{figure}[htb]
	\centering
	\subfigure[\label{LPP1}]{
		\begin{minipage}[t]{0.4\linewidth}
			\centering
			\includegraphics[width=7.2cm]{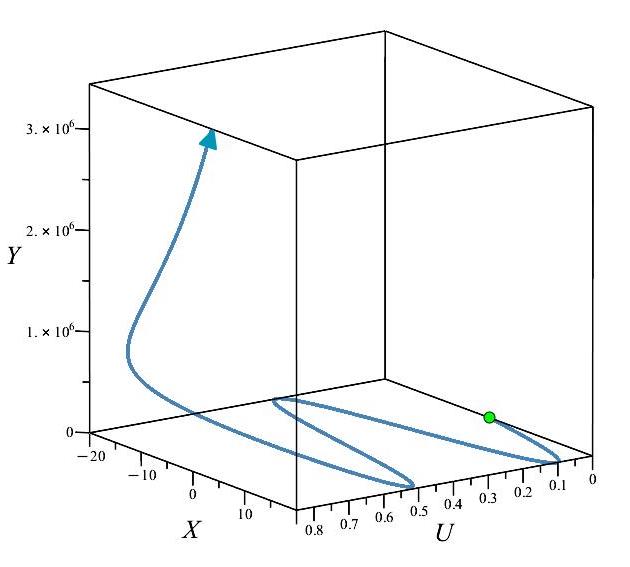}
		\end{minipage}
	}
	\subfigure[\label{LPP2}]{
		\begin{minipage}[t]{0.4\linewidth}
			\centering
			\includegraphics[width=7.2cm]{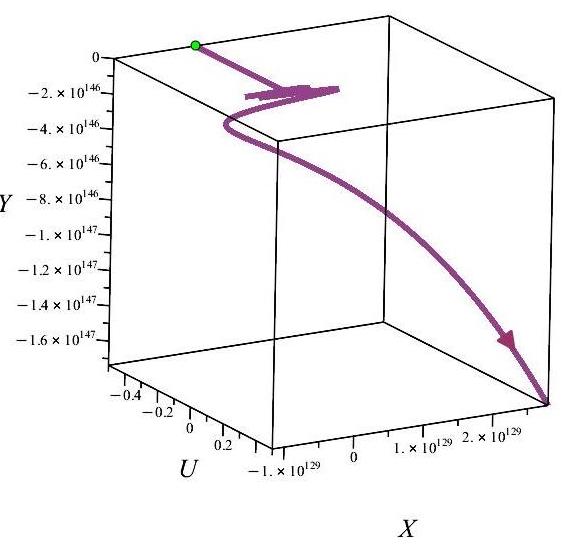}
		\end{minipage}
	}
	\caption{\textit{\small \ref{LPP1}: 
			a particle  in the LPP GW space-time \eqref{usual-LPP} of Brdicka (drawn in steel blue) oscillates. It should be compared with what happens in the rational LPP GW \eqref{LPP-pulse}, obtained by squeezing the wave as in \eqref{S-Gibbons} and drawn in dark orchid in FIG. \ref{LPP2}, for which the particle initially in rest is shaken by the GW and then escapes with straightened-out velocity due to the damping  factor $U^{-1}$ after the wave has passed.} 
	\label{LPP-plots}
	}
\end{figure}

\begin{figure}[htb]
  \centering
  \subfigure[\label{LPP1anal}]{
  \begin{minipage}[t]{0.4\linewidth}
  \centering
  \includegraphics[width=7.2cm]{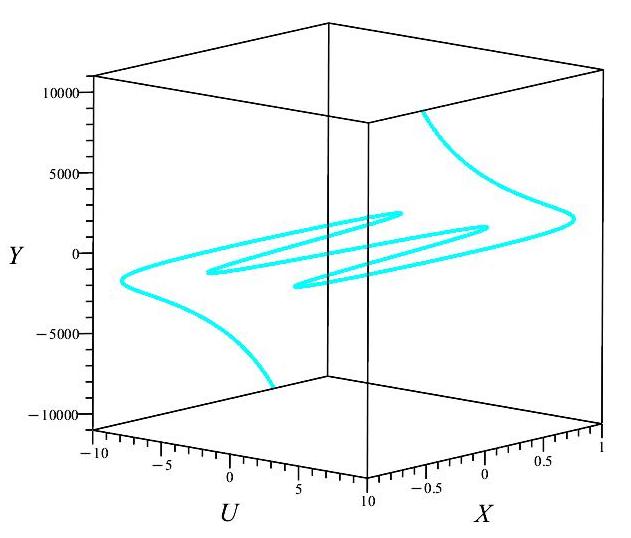}
  \end{minipage}
  }
  \subfigure[\label{LPP2anal}]{
  \begin{minipage}[t]{0.4\linewidth}
  \centering
  \includegraphics[width=7.2cm]{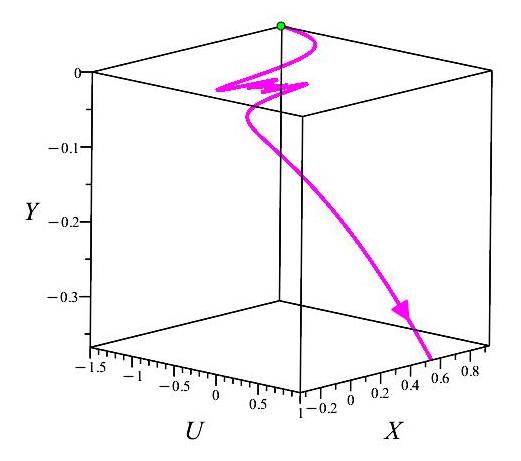}
 \end{minipage}
  }

\caption{\textit{\small  \ref{LPP1anal} shows analytically found geodesics  for the LPP (Brdicka) \eqref{usual-LPP}, and 
	\ref{LPP2anal} for the rational LPP in \eqref{Xsol-rLPP}-\eqref{Ysol-rLPP}, metric  respectively. These plots should be compared with the numerical ones in  FIG. \ref{LPP-plots}.} 
\label{analLPP-plots}}
\end{figure}

  Thus once we know the  the Killing vectors \eqref{CCKV}, we get the geodesic for free and vice versa. Below we derive the analytic formulae  by spelling out this remarkable correspondance. 
The numerical solutions shown in FIG.\ref{LPP-plots}
for the \emph{LPP} GW of Brdicka, \eqref{usual-LPP}, and for the \emph{rational LPP}, \eqref{LPP-pulse}, are matched by the analytic solutions deduced  from~\eqref{Carroll-V-L-1} and a piecewise continuous solutions deduced from \eqref{Carroll-V-L-2}, 
\begin{eqnarray}
	&&
	\widetilde{X}(\widetilde{U})=
	\left\{
	\begin{aligned}
		&\widetilde{U}\sin\widetilde{U}^{-1} \qquad \qquad \qquad \qquad
		&\widetilde{U}<0\, 
		\\
		&\widetilde{U}\cos\widetilde{U}^{-1} 
		\qquad \qquad \qquad \qquad
		&\widetilde{U}>0\,,
	\end{aligned}
	\right. 
	\label{Xsol-rLPP} 
	\\[8pt]
	&&
	\widetilde{Y}(\widetilde{U})=
	\left\{
	\begin{aligned}
		&0, \quad &\widetilde{U}\leq 0\,, \\
		&\widetilde{U}\left(\sinh\widetilde{U}^{-1}
		-\cosh\widetilde{U}^{-1}\right)\,, \quad &\widetilde{U}>0.
	\end{aligned}
	\right. \label{Ysol-rLPP}
\end{eqnarray}
These analytical solutions are plotted in FIG. \ref{analLPP-plots}.

\begin{figure}[htb]
	\centering
	\subfigure[\label{CPP1}]{
		\begin{minipage}[t]{0.4\linewidth}
			\centering
			\includegraphics[width=7.2cm]{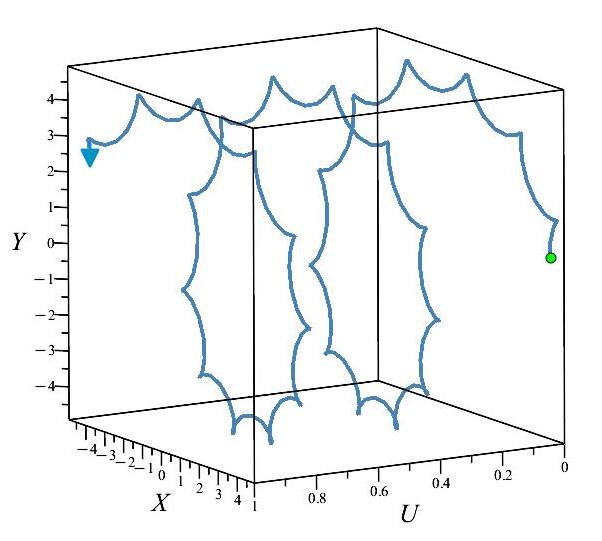}
		\end{minipage}
	}\qquad\quad
	\subfigure[\label{CPP2}]{
		\begin{minipage}[t]{0.4\linewidth}
			\centering
			\includegraphics[width=7.5cm]{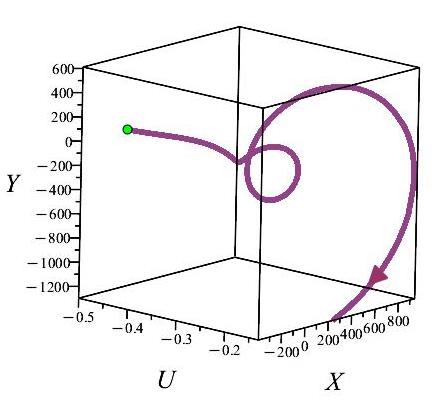}
		\end{minipage}
	}
	\\
	\caption{\textit{\small 
			\ref{CPP1}~: in the usual CPP GW (depicted in steel blue) the  particle performs a 
			``gear wheel - like'' motion.   
			\ref{CPP2}~: in the rational CPP GW (in dark orchid)  the particle which is at rest before the GW arrives escapes along an expanding screw  after the GW has passed. For large $U$ its velocity becomes approximately constant due to the damping factor $\widetilde{U}^{-4}$ in \eqref{CPP-pulse-metric}}.
		\label{CPP-plots}}
\end{figure}

The geodesics of both the CPP GW \eqref{CPP-metric} and the rational-time CPP GW \eqref{CPP-pulse-metric} perform  
screw-like motions. FIG. \ref{CPP-plots} compares these two numerically-obtained geodesics.
Eqn. \eqref{Geo-CPP-1} is an analytically found
geodesic in the \emph{CPP GW} space-time \eqref{CPP-metric} which,  choosing the parameters as  $\omega=1.5$, $\delta_1=0$, $\delta_2=0$, $\beta_1=0$ and $\beta_2=5$, matches the numerical one in FIG.\ref{CPP1}. 

\begin{figure}[htb]
	\centering
	\includegraphics[scale=.364]{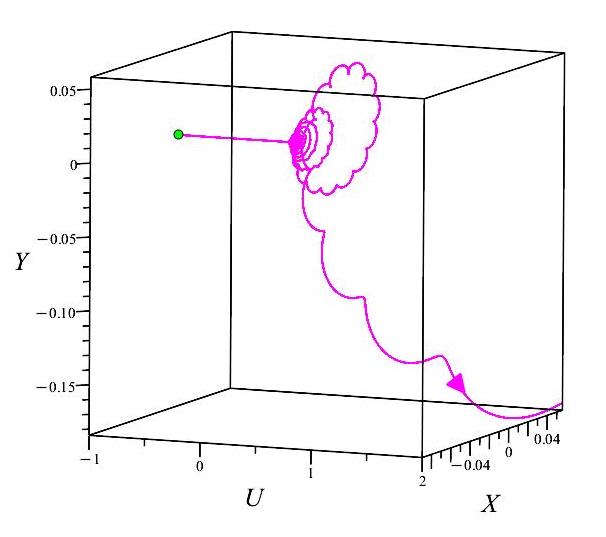}
	\caption{\textit{\small  
			The analytic rational CPP solution  \eqref{Xsol-rLPPpiecewise}-\eqref{Ysol-rLPPpiecewise}, to be compared with 
			the numerically found one in \ref{CPP2}.
		}
		\label{CPP2anal}}
\end{figure}

\begin{figure}[htbp]
	\centering
	\subfigure[\label{rCPP-vxanal}]{
		\begin{minipage}[t]{0.3\linewidth}
			\centering
			\includegraphics[width=5.8cm]{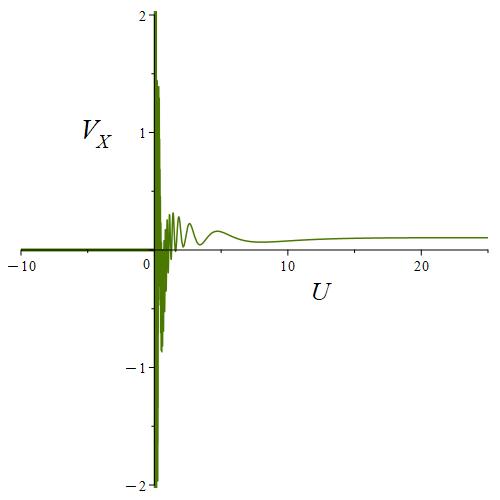}
		\end{minipage}
	}
	\subfigure[\label{rCPP-vyanal}]{
		\begin{minipage}[t]{0.3\linewidth}
			\centering
			\includegraphics[width=5.8cm]{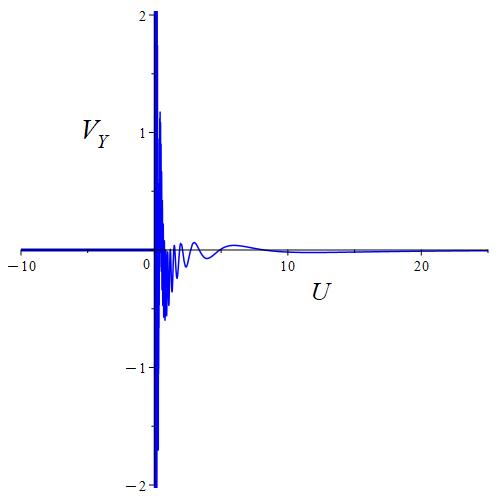}
		\end{minipage}
	}
	\\
	\caption{\textit{\small For the rational CPP GW \eqref{CPP-pulse-metric}	the velocities  become approximately
			constant after the wave  has passed due to the damping factor $\widetilde{U}^{-4}$: we get  the velocity effect \cite{Grishchuk1989,Zhang:2018srn}.}
		\label{rCPP-V-Manal}
	}
\end{figure}

The \emph{rational CPP GW} \eqref{CPP-pulse-metric} admits special piecewise solutions,
\begin{eqnarray}
	&&
	\widetilde{X}(\widetilde{U})=
	\left\{
	\begin{aligned}
		&0, \quad \widetilde{U}\leq0, 
		\\[4pt]
		&\frac{\widetilde{U}}{\omega}\mathcal{D}_{\widetilde{U}}\left(\widetilde{U}\cos\frac{\omega_-}{\widetilde{U}\;}\cdot\widetilde{U}\sin\frac{\omega}{\widetilde{U}}\right), \quad \widetilde{U}>0.
	\end{aligned}
	\right. 
	\label{Xsol-rLPPpiecewise} 
	\\[14pt]
	&&
	\widetilde{Y}(\widetilde{U})=
	\left\{
	\begin{aligned}
		&0, \quad \widetilde{U}\leq 0, 
		\\[4pt]
		&\frac{\widetilde{U}}{\omega}\mathcal{D}_{\widetilde{U}}\left(\widetilde{U}\cos\frac{\omega_-}{\widetilde{U}\;}\cdot\widetilde{U}\cos\frac{\omega}{\widetilde{U}}\right), \quad \widetilde{U}>0\,,
	\end{aligned}
	\right.
	\label{Ysol-rLPPpiecewise}
\end{eqnarray}
plotted in FIG. \ref{CPP2anal} which should be compared  with the numerical solution in FIG.\ref{CPP2}.
FIG. \ref{rCPP-V-Manal} shows the variations of the velocities in FIG. \ref{CPP2anal} on $X$ and $Y$ directions.

\section{$\Ort(2,1)$-conformally invariant gravitational waves} \label{secIV}

In the previous section we discussed  vacuum GWs that are carried into another vacuum GW by the special M\"obius  transformation \eqref{ABCD-U}-\eqref{M-trans}. In this section we consider a special vacuum GWs which are \emph{invariant}. 

We start by completing \eqref{ABCD-U} by the well-known  $\xi$-preserving conformal transformations 
of the conformal Killing equations in the free Minkowski metric in $2+1$ dimensions,  \eqref{pp-wave} with $H=0$.  We get three special transformations, namely,
\besub
\begin{align}
&\mathrm{time}\mathrm{-}\mathrm{translation}: \ U=\widetilde{U}+\epsilon, \ \ \mathbf{X}= \widetilde{\mathbf{X}}\,, \ \ V=\widetilde{V}\,, 
\label{O21-trans-1}
\\[6pt]
&\mathrm{dilatation}: 
\
U=e^{2\delta}\widetilde{U}, \ \ \mathbf{X}=e^{\delta}\widetilde{\mathbf{X}}\,, \ \ V=\widetilde{V}, 
\label{O21-trans-2}
\\[4pt]
&\mathrm{special}\ \mathrm{conformal}\ \mathrm{transformation}: 
\nn \\[4pt]
& \qquad U=\frac{\widetilde{U}}{1+\kappa\,\widetilde{U}}\,, \;\;\;\;
 \mathbf{X}=\frac{\widetilde{\mathbf{X}}}{1+\kappa\,\widetilde{U}}\,, 
\;\;\;\;
 V=\widetilde{V}+\frac{\kappa}{2(1+\kappa\,\widetilde{U})}\widetilde{\mathbf{X}}^2\,,
\label{O21-trans-3}
\end{align}
\label{finiteO21tr}
\esub
where $\epsilon$, $\delta$ and $\kappa$ are arbitrary real constants. The corresponding infinitesimal generators,
\besub
\begin{align}
&\mathrm{time}\mathrm{-}\mathrm{translation}: \
&\widehat{E}=\partial_{U},
\label{T-trans}
\\
&\mathrm{dilation}: \ &\widehat{D}=U\partial_U+\frac{1}{2}\left(X\partial_X+Y\partial_Y\right),
\label{Dilation}
\\
&\mathrm{expansion}: \ &\widehat{K}=U^2\partial_U
+U\big(X\partial_X+Y\partial_Y\big)-\frac{1}{2}(X^2+Y^2)\partial_V\,,
\label{Expansion}
\end{align}
\label{O21tr}
\esub
span an $\ort(2,1)$ algebra,
\begin{eqnarray}
[\widehat{D},\widehat{E}]=-\widehat{E}, \quad [\widehat{D},\widehat{K}]=\widehat{K}, \quad [\widehat{E},\widehat{K}]=2\widehat{D}\,
\end{eqnarray}
which generate an $\Ort(2,1)$ conformal group.
\goodbreak

\medskip
Systems with $\Ort(2,1)$ 
symmetry were considered in various physical instances:
\begin{itemize}
\item For a free particle \cite{SchrSymm,DBKP,DGH91} or in Chern-Simons field theory  \cite{JackiwPiCS,DHP1,DHP2}  it extends the Galilei to the Schr\"odinger algebra \cite{SchrSymm}. All Schr\"odinger-symmetric systems are derived, in $d\geq 3$ space dimensions,  from the vanishing of the Weyl \cite{DHP2} or in in $d= 1$ from that of the Cotton tensor \cite{Zurab}, respectively.

\item 
An inverse-square potential could be added \cite{DBKP,Jacobi,SchrSymm,DHP2,DHP1,JackiwPiCS,AFF,ScalingZCEH,Zurab}. Applications include  the interaction of a polar molecule with an electron \cite{Camblong:2001zt,Moroz:2009nm} (which will be discussed further in subsec.\ref{PolarSec}), the Efimov effect \cite{Efimov:1973awb,Moroz:2009nm}, near-horizon fields of black holes \cite{Camblong:2003mz,Claus,Moroz:2009nm} and the vacuum AdS/CFT correspondence \cite{Gubser:1998bc,Moroz:2009nm,Witten:1998qj} ;

\item A Dirac-monopole and a magnetic vortex \cite{Jackiw:1980mm,Jackiw:1989qp}\,.
\end{itemize}

Hence we focus our attention at vacuum gravitational waves with $\Ort(2,1)$ symmetry. For symplicity,  we focus our investigations to the planar case with coordinates $X,\,Y$.
Substituting the three vectors in 
\eqref{T-trans}-\eqref{Expansion} into the conformal Killing equation \eqref{L-D} for the Brinkmann metric \eqref{pp-wave} leaves us  with,
\besub
\begin{align}
&\mathrm{time}\mathrm{-}\mathrm{translation}: \ H_{,U}=0\,,
\label{timetrsym}
\\
&\mathrm{dilatations}:
\  UH_{,U}+XH_{,X}+YH_{,Y}+2H=0\,,
\label{dilatsym}
\\
&\mathrm{special}\ \mathrm{conformal}\ \mathrm{transformation}: \
2UH_{,U}+XH_{,X}+YH_{,Y}+2H=0\,.
\label{expansym}
\end{align}
\label{three-eqs}
\esub
Note that  \eqref{dilatsym} and \eqref{expansym}  differ only in the  coefficients of their first terms -- which involves the generator of time translation symmetry, \eqref{timetrsym}.

Solving these equations with the vacuum condition \eqref{vacuum condition} yields, for an exact plane wave, the line element,
\begin{eqnarray}
\bigbox{
ds_{O21}^2=dX^2+dY^2+2dUdV-2\left(\frac{C_1(X^2-Y^2)+2C_2XY}{R^4}\right)dU^2\,,}
\label{O21-pp-wave+}
\end{eqnarray}
where $R^2=X^2+Y^2$; $C_1$ and $C_2$ are arbitrary constants. The proof follows at once from that
 dilatation symmetry \eqref{dilatsym}, combined with time-translation-invariance \eqref{timetrsym} imply indeed, by Euler's formula, that the potential is homogeneous of order $(-2)$. 

The  potential \eqref{O21-pp-wave+} breaks   the rotational symmetry, however still allows for the conformal $\Ort(2,1)$ symmetry of the  inverse-square potential \cite{AFF,DBKP,DGH91} to the anisotropic case. 
 It should  be compared to the statement \cite{DHP2,Zurab} which says that the profiles of the
only Bargmann manifolds with \emph{Schr\"odinger} symmetry  
correspond, in  3+1 dimensions, to an (i) isotropic oscillator, to an (ii) inverse-square potential with constant coefficient, to a (iii) uniform force field. 

 The special GW \eqref{O21-pp-wave+} satisfies, for  
an arbitrary linear combination of $\widehat{E}, \,\widehat{D},\, \widehat{K}$ in \eqref{T-trans}-\eqref{Expansion}, the conformal Killing equations \eqref{L-D} with,
\beq
W=a\widehat{E}+b\widehat{D}+c\widehat{K}=
(a+bU+cU^2)\p_{U}-c\frac{1}{2}\mathbf{X}^2\p_{V}+(cU+\frac{b}{2})\mathbf{X}\cdot\p_{\mathbf{X}}\,, 
\label{combin-KVs}
\eeq
where  $a$, $b$ and $c$ are arbitrary constants.
By integrating the $U$ component of \eqref{combin-KVs}, the associated SKV reduces to the M\"obius-redefined time \eqref{ABCD-U} with $\mathbf{X},\, A$, etc replaced by, $\tilde{\bX}$ and
by,
\beq\tilde{A}=-\left[\frac{b}{2}+\frac{\sqrt{4ac-b^2}}{2\tan\left(\frac{\eta\sqrt{4ac-b^2}}{2}\right)}\right],
\;\;
\tilde{B}=-c,
\;\; 
\tilde{C}=a,
\;\;
\tilde{D}=-\left[\frac{\sqrt{4ac-b^2}}{2\tan\left(\frac{\eta\sqrt{4ac-b^2}}{2}\right)}\right]\,,\quad
\label{Combin-O21-func}
\eeq
where $\eta$ is the parameter of the integral curve.
In conclusion, the special gravitational wave  \eqref{O21-pp-wave+} is form-invariant under  the SMCT \eqref{M-trans}.

The metric \eqref{O21-pp-wave+} is conveniently presented in cylindrical coordinates $(R,\theta)$,
\begin{eqnarray}
ds_{O21}^2=
dR^2+R^2d\theta^2+2dUdV-2\left(\frac{C_1\cos2\theta+C_2\sin2\theta}{R^2}\right)dU^2\,,
\label{nosympot}
\end{eqnarray} 
 reminiscent of  the  potential  for the  interaction between a \emph{polar molecule and an electron}   \cite{Camblong:2001zt,Camblong:2003mz},
 \beq
H\equiv H(r,\theta)=\frac{C\cos\theta}{r^2}\,,
\label{polmolpot}
\eeq
where  the constant $C$ is proportional to the product of the electric charge and  the dipole momentum, and
$\theta$ is the polar angle in the $X-Y$ plane. 

\section{A molecular physics-inspired spacetime}\label{PolarSec}

In this section, we study a vacuum spacetime  inspired by polar molecules represented by the anisotropic inverse-square potential \cite{Camblong:2001zt,Camblong:2003mz},
\beq
H = \frac{C_1\cos2\theta+C_2\sin2\theta}{R^2}\,,
\label{polpot}
\eeq
 where $C_1$ and $C_2$ are real constants, cf. \eqref{nosympot}.
Postponing the $3$-dimensional problem to further study, we limit our attention at the plane. For simplicity, we put also the NR mass $M=1$.
The conformal Killing vectors in \eqref{T-trans}-\eqref{Dilation}-\eqref{Expansion} preserve the vertical vector $\xi=\p_V$ and therefore project to conformal symmetries of the underlying non-relativistic system  providing us with three conserved quantities \cite{SchrSymm,AFF,DBKP,Jacobi},
\besub
\begin{align}
\widehat{E}\quad\rightarrow \quad &\mathcal{E}=\frac{\mathbf{P}^2}{2}+\frac{C_1\cos2\theta+C_2\sin2\theta}{R^2}\,, 
\label{cenergy}
\\[4pt]
\widehat{D} \quad \rightarrow \quad &\mathcal{D}=2\mathcal{E}U-\mathbf{P}\cdot\mathbf{X}\,,
\label{cdilat} 
\\[4pt]
\widehat{K} \quad \rightarrow \quad &\mathcal{K}=
-\mathcal{E}U^2+\mathcal{D}U+\frac{1}{2}R^2\,.
\label{cexpan}
\end{align}
 \label{o21CQ}
\esub
To explain  in simple terms what happens, consider first dilations, \eqref{Dilation}, which leave the  Lagrange density $L_0dU$ of a free NR particle invariant provided the time scales with the square of the factor as the position  does \cite{SchrSymm}. Then adding a potential $H$ changes the Lagrange density by $- HdU$, which is also invariant if $H$ is inverse-square in the radius \cite{AFF,DBKP,ScalingZCEH}. 

However dilations act only on the radial variable, therefore the potential \eqref{polpot} is left invariant. Then an easy calculation shows that the two other transformations in \eqref{O21tr} remain also unbroken.
Remarkably, the associated ``Noether" quantities were found by Jacobi \cite{Jacobi} \dots 60 years \emph{before} Emmy Noether was  born~! 

The Casimir operator of  $\Ort(2,1)$ is,
\begin{eqnarray}
\mathcal{C}^2=\mathcal{R}^2-\mathcal{G}_-^2-\mathcal{G}_+^2, 
\label{CC}
\end{eqnarray}
where
\begin{eqnarray}
\mathcal{R}=\frac{1}{2}\left(\frac{1}{\tau}\widehat{K}+\tau \widehat{E}\right), \quad \mathcal{G}_{-}=\frac{1}{2}\left(\frac{1}{\tau}\widehat{K}-\tau \widehat{E}\right), \quad \mathcal{G}_{+}=\widehat{D}\, 
\label{RG1G2}
\end{eqnarray}
generate a compact $\SO(2)$ group of rotations, augmented with two non-compact two dimensional boosts. Here $\tau$  is a positive fixed parameter with the dimension of time. See  Ref. \cite{Jackiw:1989qp} for details. 
The Casimir operator can also be written as,
\begin{eqnarray}
\mathcal{C}^2&=&{J}^2
+2\left(C_1\cos2\theta + C_2\sin2\theta\right), 
\label{Casimir CQ}
\end{eqnarray}
where ${J} =  \mathbf{R}\times\mathbf{V}$ is the orbital angular momentum. ( The angular momentum in 2 dimensions is just a scalar, namely the 3rd component of the 3-dimensionalone, $J_z$. The conserved quantity generated by translations along the $V$ coordinate and interpreted as the mass of the underlying non-relativistic system \cite{DBKP,Eisenhart,DGH91} was scaled to unity).

A lightlike particle in the special GW  background \eqref{O21-pp-wave+} (viewed, in the Bargmann framework,  as a massive non-relativistic particle in one dimension less) moves along null geodesics. In cylindrical coordinates,
\begin{eqnarray}
&&\frac{\;d^2R}{dU^2}-R\left(\frac{d\theta}{dU}\right)^2
-\frac{2\left[C_1\cos2\theta+C_2\sin2\theta\right]}{R^3}=0\,, 
\label{geo-eq-1}
\\[8pt]
&&\frac{\;d^2\theta}{dU^2}+\frac{2}{R}\frac{dR}{dU}\frac{d\theta}{dU}
-\frac{2\left[C_1\sin2\theta-C_2\cos2\theta\right]}{R^4}=0\,. 
\label{geo-eq-2}
\end{eqnarray}


Let us assume, for simplicity, that $C_1=0$ so that the planar metric \eqref{nosympot} has only one polarization state, 
\begin{eqnarray}
\bigbox{
ds^2=
dR^2+R^2d\theta^2+2dUdV-2\left(\frac{C_2\sin2\theta}{R^2}\right)\,dU^2\,, }
\label{C10pot}
\end{eqnarray}

For $C_2=0$  we get Minkowski-space which has no interest for us. Then
$C_2>0$ can be achieved by shifting $\theta$. Henceforth we set $C_2=1$. 

\begin{figure}[h] 
\hskip-6mm\includegraphics[scale=.39]{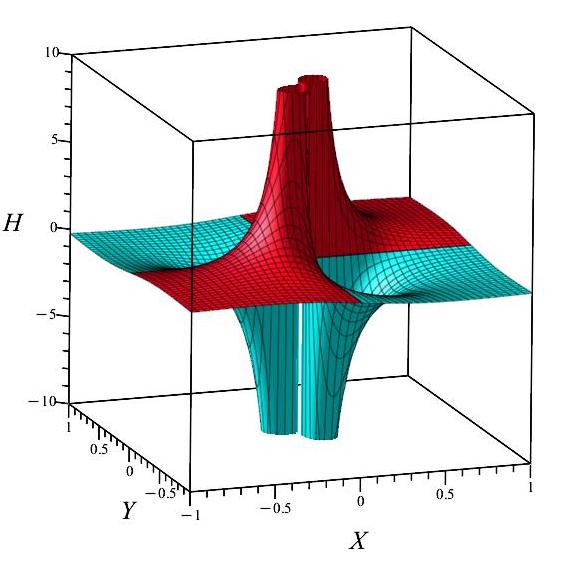}\,
\includegraphics[scale=.41]{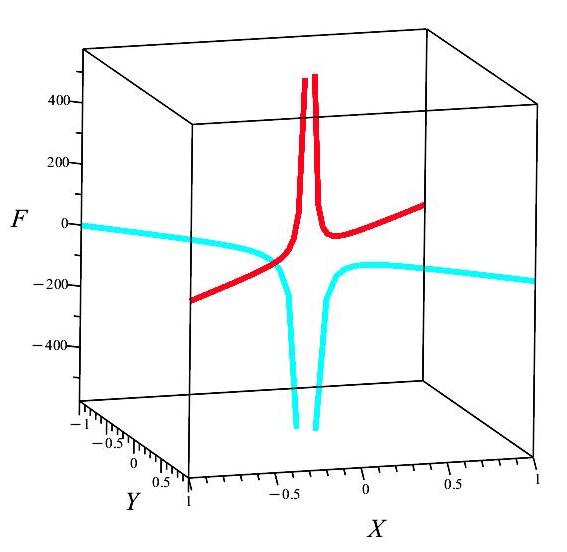}
\\
{}\vskip-3mm
(a)\hskip82mm (b)
\vskip-3mm\caption{\textit{\small  \ref{nosympotplot}(a): the potential 
 \eqref{C10pot} alternates between \red{\bf repulsive (NE-SW)} and \cyan{\bf attractive (NW-SE)}, 
  changing sign at every quadrant. The apparent doubling of the ``chimneys and sinks'' in FIG.\ref{nosympotplot}a are computer artifacts as confirmed by FIG.\ref{nosympotplot}b~: the only singularity is at the origin.}
 \label{nosympotplot}
}
\end{figure}

\begin{figure}[h]
\includegraphics[scale=.38]{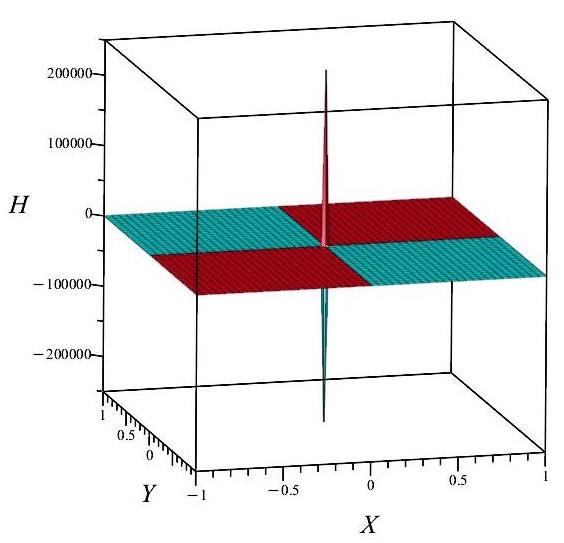}\vskip-5mm\caption{\textit{\small A long-distance view of the wave \eqref{C10pot} shows a ``spike'' whose sign alternates at every quarter of the circle. 
}}
\label{spike}
\end{figure}

The metric \eqref{C10pot} is the ``Bargmannian'' form \cite{Eisenhart,DBKP,DGH91} of the anisotropic version of a NR particle in an inverse-square potential
\beq
H(R,\theta) = \frac{\sin2\theta}{R^2}\,,
\label{effpot}
\eeq
shown in FIG.\ref{nosympotplot}. Its anisotropy is manifest by realizing that for fixed  $R=R_0$, $H(R,\theta)$ is proportional to $\sin 2\theta$. 
 A long-distance view is shown in  FIG.\ref{spike}.
 
\begin{figure}[h]
\includegraphics[scale=.44]{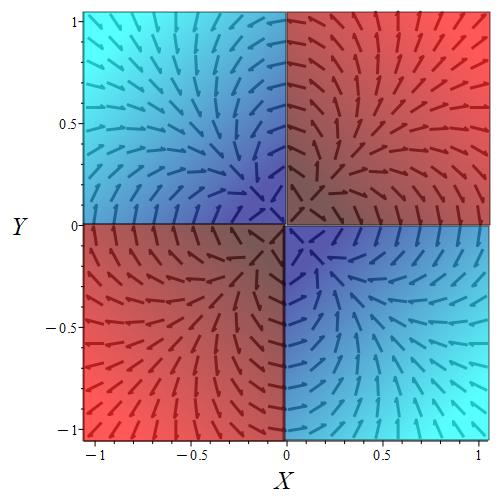}
\\
\vskip-6mm\caption{\textit{\small  The force  $-\grad{H}$ 
 alternates  at every quarter-of-circle between \red{\bf  repulsive (NE - SW)} and \cyan{\bf attractive (NW - SE)} regions. The force is \red{maximally repulsive} along the ``crests'' at \red{$\pi/4$} and  
  \red{$5\pi/4$} and \cyan{maximally attractive} in the \cyan{``valley bottoms''} at \cyan{$3\pi/4$} and \cyan{$7\pi/4$}, respectively.} 
\label{forceplot}
}
\end{figure}

The nature of the potential \eqref{C10pot} 
is determined by the sign of the coefficient of $dU^2$ ---
the potential of  the underlying non-relativistic dynamics
\cite{DBKP,DGH91} --- which alternates at every quadrant.
Its behavior is conveniently studied by plotting the force, FIG.\ref{forceplot}:
It is
\red{\bf repulsive} for \red{$0 < \theta < \pi/2$} and for \red{$\pi < \theta < 3\pi/2$}, and \cyan{\bf attractive} for 
\cyan{$\pi/2 < \theta < \pi$} and for \cyan{$3\pi/2 < \theta < 2\pi$}. The force is maximal
 on the separation ``crosslines''  at $\theta = k\pi/2,\, k=0, 1, 2, 3$, where the  repulsive potential becomes  attractive and vice versa, cf. \eqref{effpot}. It is obviously symmetric w.r.t.  $\theta \to \theta+\pi$.
   
A qualitative insight into the possible motions can be obtained by using the conformal $\ort(2,1)$ symmetry. For simplicity we restict our attention at  what happens to a particle that we simply put at $U=U_0$ to some position $(R_0, \theta_0)$ 
with vanishing initial velocity. Then the conserved quantities \eqref{o21CQ} generated by $\ort(2,1)$ reduce, putting $M=1,\, C_1=0,\, C_2=1$ for simplicity, to
\besub
\begin{align}
&\mathcal{E}_0=\frac{\sin2\theta_0}{R_0^2}\,, 
\label{cenergy-0}
\\[4pt]
&\mathcal{D}_0=2\mathcal{E}_0U_0\,,
\label{cdilat-0} 
\\[4pt]
&\mathcal{K}_0=
-\mathcal{E}_0U_0^2+\mathcal{D}_0U_0+\frac{1}{2}R_0^2\,.
\label{cexpan-0}
\end{align}
\label{o21CQ0}
\esub 
From \eqref{cenergy-0} we deduce that the conserved energy, which is initially just the potential, may be \red{\bf positive}, \cyan{\bf negative} or {\bf zero}, 
corresponding to the repulsive or attractive quadrant or to the separation line between them, as depicted in FIG.s \ref{nosympotplot} and \ref{forceplot}.

\benu

\item In the
\red{repulsive quadrants} $0<\theta<{\pi}/{2}$ or $\pi<\theta<{3\pi}/{2}$ 
the energy is positive,
\begin{eqnarray}
\mathcal{E}=\mathcal{E}_0 =\frac{\mathbf{P}^2}{2}+\frac{\sin2\theta}{R^2}>0\,
\Rightarrow
\frac{\mathbf{P}^2}{2}>\left|\frac{\sin2\theta}{R^2}\right|.
\end{eqnarray}
Thus  the motion is outgoing.
When the particle crosses the separation line and enters into the attractive area, the absolute value of negative potential is less than that of  the initial potential: the particle will be pushed out to infinity.

\item
In the \cyan{attractive quadrants}, ${\pi}/{2}<\theta<\pi$ or ${3\pi}/{2}<\theta<2\pi$ the energy is negative,
\begin{eqnarray}
\mathcal{E}=\mathcal{E}_0 =\frac{\mathbf{P}^2}{2}+\frac{\sin2\theta}{R^2}<0\,
\Rightarrow
\frac{\mathbf{P}^2}{2}<\left|\frac{\sin2\theta}{R^2}\right|.
\end{eqnarray}
Thus the kinetic energy is dominated by the potential energy, and we get incoming motion with the particle falling into the hole.

\item
An intermediate behaviour is observed for vanishing energy
when the initial position is on one of the a \textbf{separation line} between repulsive and attractive quadrants, i.e., for $\theta_k= \, k\frac{\pi}{2},\, k=0,1,2,3$~: by \eqref{cenergy-0} and \eqref{cdilat-0} we have, 
\begin{eqnarray}
\mathcal{E}=\mathcal{E}_0=0 
\aand
\mathcal{D}=\mathcal{D}_0=-\mathbf{P}\cdot\mathbf{X}=0. 
\label{D_0K0}
\end{eqnarray}
so that \eqref{cexpan-0} implies that
\beq
R = R_0 = \const  \aand \mathbf{P} \perp \bX\,.
\label{Rconst}
\eeq
In conclusion,  a particle put  on the ``rim''  will follow a circular trajectory inside the attractive region. 
Moreover, the vanishing of the energy,
\beq
2\mathcal{E}_0 =
\mathbf{P}^2+2\frac{\sin2\theta}{R^2}=0,
\label{zeroenergy}
\eeq
 implies that the particle oscillates between the ``rims" of the attractive quadrants,
 \beq
\frac{\pi}{2}\leq\theta\leq\pi \oor \frac{3\pi}{2}\leq\theta\leq2\pi\,.
\label{0oscill}
\eeq 
\eenu
 
\begin{figure}[h]
\includegraphics[scale=.3]{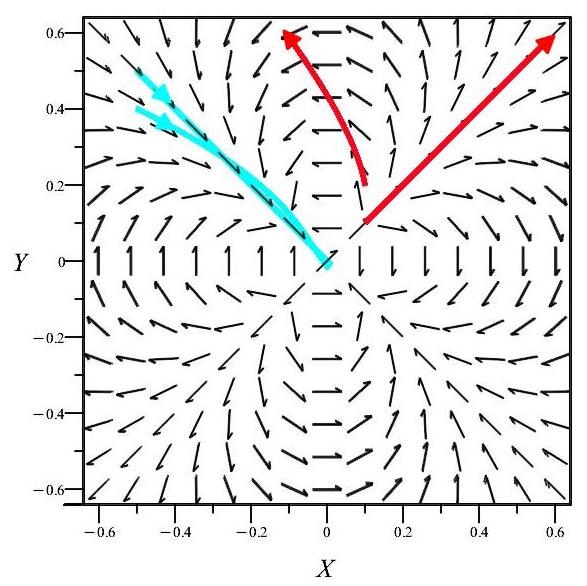}\vskip-3mm\caption{\textit{\small Particles which start in the \red{repulsive} zone are pushed to infinity both in the repulsive quadrant, and, after crossing over, also in the \cyan{attractive} quadrant. Particles which start from the \cyan{attractive} zone are in turn sucked into the hole.
This behavior corresponds to the sign of the non-relativistic energy \eqref{cenergy-0}. 
}}
\label{out-in-XY}
\end{figure}

\begin{figure}[htbp]
  \centering
  \subfigure[\label{Ia}]{
  \begin{minipage}[t]{0.4\linewidth}
  \centering
  \includegraphics[width=6cm]{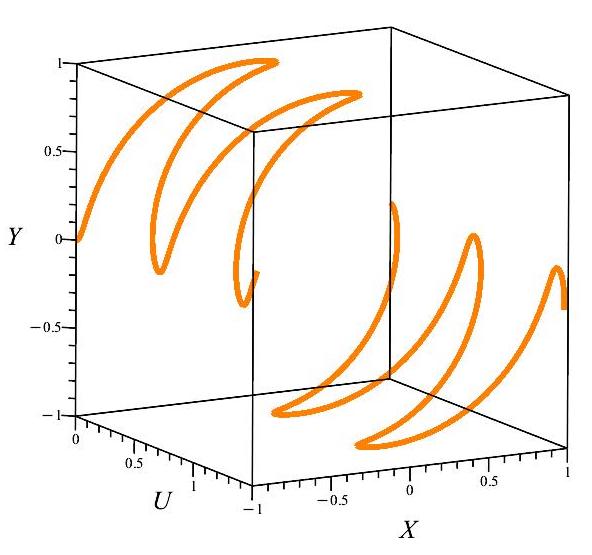}
  \end{minipage}
  }
  \subfigure[\label{Ib}]{
  \begin{minipage}[t]{0.4\linewidth}
  \centering
  \includegraphics[width=5.3cm]{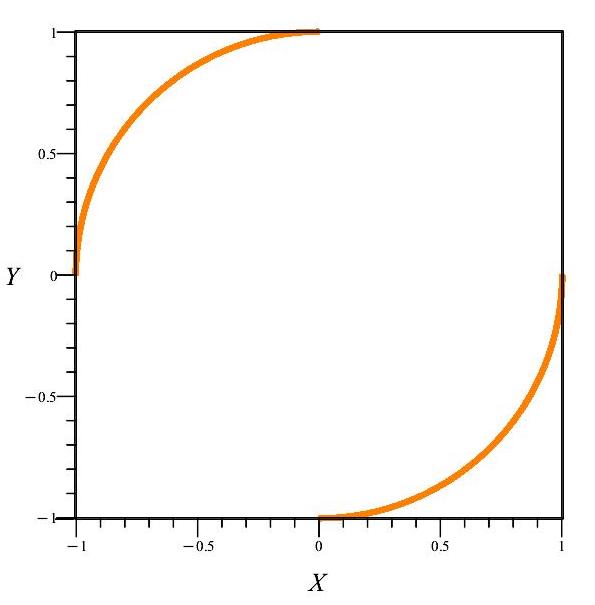}
 \end{minipage}
 }
 \\
 \caption{\textit{\small Numerically obtained  periodic trajectories \ref{Ia} in the $\ort(2,1)$ symmetric but non-isotropic gravitational wave \eqref{C10pot}.  
  \ref{Ib} shows their projections onto the $X-Y$ plane, as seen also in FIG.\ref{Allmotions}. 
 The   curves show two particles which start  from $(1,0)$ resp. at $(-1,0)$ with zero initial velocity. 
The trajectories oscillate along  quarters-of-a-circle.}
\label{num-sol-1}
  }
\end{figure}

\begin{figure}[h]
\includegraphics[scale=.4]{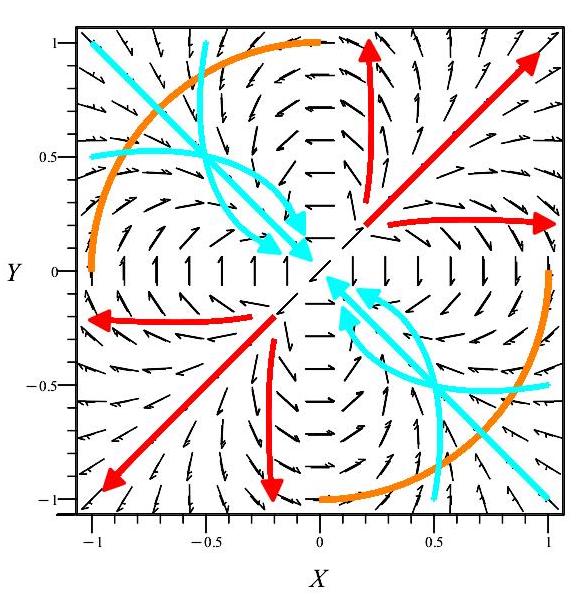}\vskip-3mm\caption{\textit{\small in the \magenta{\bf NE} and \magenta{\bf SW} quadrants the particle is pushed outwards to infinity whereas it is sucked into the  origin in the  \cyan{\bf NW} and \cyan{\bf SE} quadrants.  \orange{\bf Bounded} zero-energy motions arise which oscillate in  the  attractive quadrant between the  separation lines of the attractive and repulsives zones.
}}
\label{Allmotions}
\end{figure}

Numerical investigations indicate that the eqns \eqref{geo-eq-1}-\eqref{geo-eq-2} admit 
all three types of outgoing/infalling/bounded solutions. The first two are shown in  FIG.\ref{out-in-XY},
and the circularly oscillating  one in FIG.\ref{num-sol-1}.
The general behavior is summarised in FIG.\ref{Allmotions}.

Analytic solutions can be found also. 

We  first inquire about radial motions.
Putting $\theta=\theta_0=\const$ into \eqref{geo-eq-1}-\eqref{geo-eq-2} yields,
\begin{eqnarray}
\frac{\;d^2R}{dU^2}-\frac{2\sin2\theta_0}{R^3}=0\, 
\aand
\frac{2\cos2\theta_0}{R^4}=0\,. 
\nn
\end{eqnarray}
The 2nd eqn  
implies, 
\beq
\theta_0=(2\ell+1)\frac{\pi}{4}\,,\qquad \ell=0,\, 1,\,2,\,3\,,
\label{diagmot}
\eeq 
leaving us with the familiar inverse-square-potential equation,
\beq
\frac{d^2R}{dU^2} = {\pm}\frac{2}{R^3}\,,
\label{radRU}
\eeq
where the sign is positive in the repulsive, $\ell = 0,\, 2$ case and is negative in the attractive, $\ell=1,\, 3$  one. Thus for $\ell$ pair 
the particle is expulsed  to infinity  along the ``crest'', and for $\ell$ odd it is sucked into the origin along the ``valley bottom'' which correspond to the {maximally repulsive} or {maximally attractive}  directions in FIGs. \ref{nosympotplot} and \ref{forceplot}. 
\bmagenta
For motion along the diagonals the solution is \cite{AFF,DBKP,DHP2,ScalingZCEH}, 
\beq
R(U)=\sqrt{(V_0U+R_0)^2\pm\frac{2U^2}{R_0^2}}\,,
\label{RU}
\eeq
where $R_0>0$ and $V_0$ are the initial position and velocity at $U=0$, respectively. We choose $V_0=0$ for simplicity.
Then starting in the \magenta{\bf repulsive quadrants} with 
 $\theta=\pi/4$ or $\theta=5\pi/4$ we have the \magenta{plus} sign and 
\beq
R(U) \geq 
\sqrt{R_0^2+\frac{2U^2}{R_0^2}}
\geq R_0
\label{expulsion}
\eeq
increasing with $U$: the particle is expelled. 

In the \cyan{\bf attractive quadrants} with 
 $\theta=3\pi/4$ or $\theta=7\pi/4$ we have the \cyan{\bf minus} sign
and the motion is directed towards the origin~: 
\beq
R(U)= \sqrt{R_0^2-\frac{2U^2}{R_0^2}}
\leq R_0\,,
\label{fallingin}
\eeq
which says that the particle moves inwards, however after the critical value 
\beq
U_{crit}=\frac{R_0^2}{\sqrt{2}}
\label{Ucrit}
\eeq
$R(U)$ would become imaginary, indicating that the particle has fallen into the hole.

\emagenta

\begin{figure}[h]
\includegraphics[scale=.3]{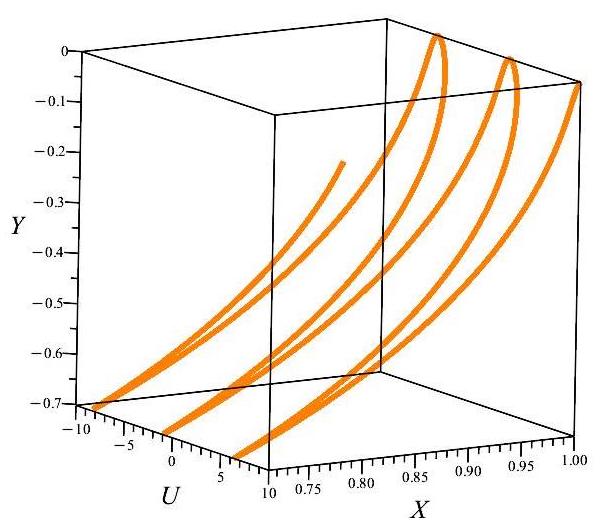}
\\
\vskip-3mm\caption{\textit{\small  The analytic solution  obtained in terms of elliptic integrals describes periodic motion along a circular arc confined into the attractive quadrant $\pi/2<\theta <\pi$, consistently with the numerical solutions in FIG.s \ref{num-sol-1} and \ref{num-sol-2}.
}
\label{analperplot}
}
\end{figure}

The equations \eqref{geo-eq-1}-\eqref{geo-eq-2} admit  also exact \emph{circular, analytic solutions}.   
 Let us indeed fix the radius, $R(U)=R_0=\const$ which reduces  \eqref{geo-eq-1}-\eqref{geo-eq-2} to,
\beq
\left(\frac{d\theta}{dU}\right)^2
+\frac{2}{R_0^4}\,\sin2\theta=0\,, 
\quad
\frac{\,d^2\theta}{dU^2}+\frac{2}{R_0^4}\,\cos2\theta=0\,. 
\label{C10eqs1}
\eeq
Deriving the first eqn. by $U$  we get
$
\frac{d\theta}{dU}\left(\frac{\,d^2\theta}{dU^2}
+\frac{2C_2}{R_0^4}\,\cos2\theta\right)=0,
$
which is an identity when the 2nd equation is satisfied.
The first equation in \eqref{C10eqs1} then implies that
\beq
\frac{d\theta}{dU}=
\big(\frac{2}{R_0^2}\big)^{1/2}\,\sqrt{-\sin2\theta}=0\,,
\label{circulareq}
\eeq
which admits real solutions when the sin is negative i.e. in 
the quadrants $\pi/2\leq \theta \leq \pi$ and $3\pi/2\leq \theta \leq 2\pi$
and is then solved in terms of  elliptic integrals  \cite{EllipticInt}, 
\begin{eqnarray}
\theta(U)=-\frac{1}{2}
\arcsin\left\{\mathrm{JacobiCN}^2\left[\frac{2}{R_0^2}(U+D),\frac{\sqrt{2}}{2}\right]\right\}\,,
\label{boundJacobi}
\end{eqnarray}
where $D$ is an integration constant. This formula can also be verified directly and is plotted in FIG.\ref{analperplot} (to be compared with the numerical solution in FIG.\ref{num-sol-1}).

This solution has zero-energy. Conversely \cite{Sundaram},  for vanishing energy ${\cal E}=0$  the conserved quantity generated by dilations, \eqref{cdilat-0} implies  $R=R_0=\const$, \eqref{Rconst}. Then \eqref{cenergy} becomes \eqref{zeroenergy} which for  $R=R_0$ is \eqref{circulareq} that we have just solved. In conclusion, the $\ort(2,1)$ symmetry implies, for zero energy, motion on (part of) a circle.  

\begin{figure}[htbp]
  \centering
  \subfigure[\label{IIa}]{
  \begin{minipage}[t]{0.3\linewidth}
  \centering
  \includegraphics[width=4.8cm]{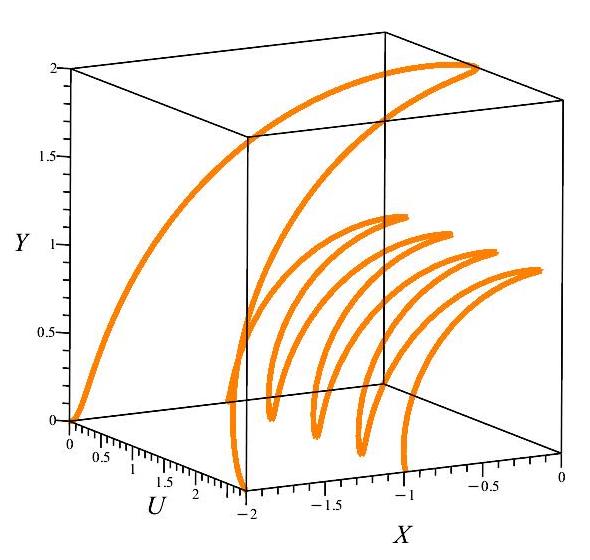}
  \end{minipage}
  }
  \subfigure[\label{IIb}]{
  \begin{minipage}[t]{0.3\linewidth}
  \centering
  \includegraphics[width=4.3cm]{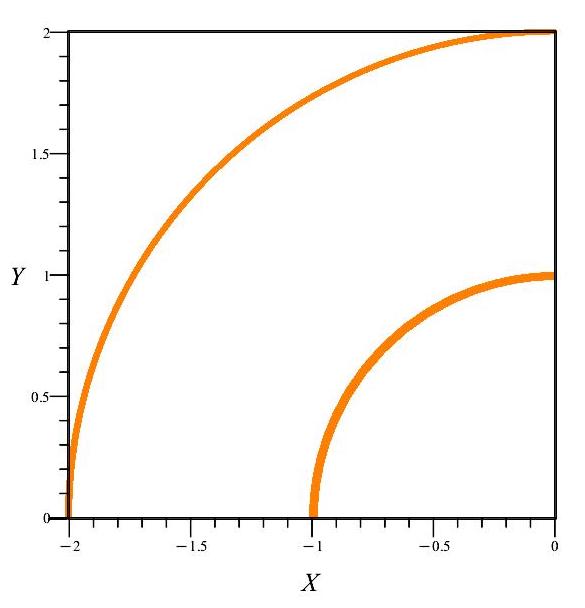}
  \end{minipage}
  }
  \subfigure[\label{IIc}]{
  \begin{minipage}[t]{0.3\linewidth}
  \centering
  \includegraphics[width=4.9cm]{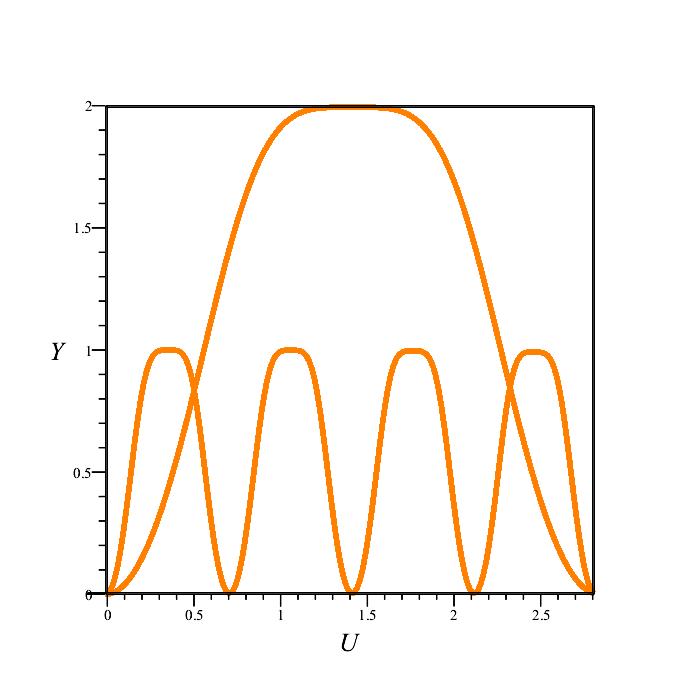}
  \end{minipage}
  }
  \\
\caption{\textit{\small FIG. \ref{IIa} shows 
 the trajectories of two  particles initially at rest on the separation line of the repulsive and attractive quadrants at \orange{\bf $(\bf{1,0})$} and at \orange{\bf $(\bf{2,0})$}, respectively.
 The  projections  in FIG. \ref{IIb} into the $X-Y$ plane follow quarter-of-circle arcs with radiuses $R_0=1$ and $R_0=2$. The projection into the $Y-U$ plane in FIG. \ref{IIc}, shows that the period for $R_0=2$ 
 is four times that for $R_0=1$, consistently with \eqref{period}.}  
 \label{num-sol-2}
  }
\end{figure}

Restoring the radius in the equations shows 
that the period increases proportionally to
the its square,  $R_0^2$,
\begin{eqnarray}
\Delta U=R_0^2\frac{4K}{\sqrt{2C_2}} \;\Rightarrow\; \Delta U
 \propto R_0^2\,,
\label{period}
\end{eqnarray}
as seen in FIG. \ref{num-sol-2}. 
Inserting  $\theta(U)$  from \eqref{boundJacobi}
into the conserved Casimir  \eqref{Casimir CQ} we get, for $C_1=0$,
\begin{eqnarray}
J^2=2\,\mathrm{JacobiCN}^2\left[\frac{2}{R_0^2}(U+C),\frac{\sqrt{2}}{2}\right]+\const 
\label{Jsquare}
\end{eqnarray}

\begin{figure}[h]
\includegraphics[scale=.38]{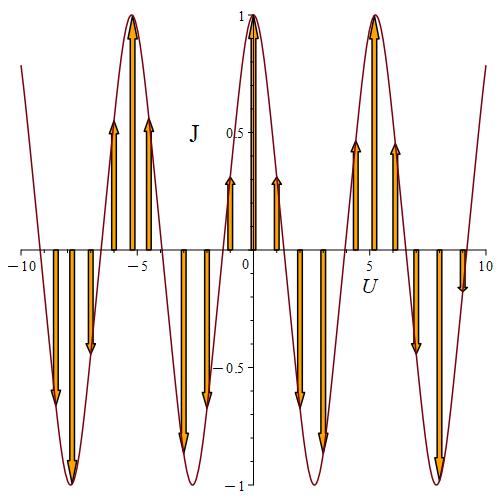}\vskip-3mm
\caption{\textit{\small In the anisotropic metric \eqref{C10pot} in FIG.\ref{nosympotplot}, the orbital angular momentum ${J}$ in \eqref{oscJvector} is 
 not conserved: for the circular periodic motion found for zero energy, for example, its
length oscillates.  The direction  of the  oscillations seen in FIGs.\ref{num-sol-1} and \ref{num-sol-2}  corresponds to the sign of the angular momentum, with the turning point corresponding to the zeros of the angular momentum. 
}}
\label{Joscifig}
\end{figure}

On the other hand, the angular momentum for  \eqref{boundJacobi} is
\begin{eqnarray}
{J} = \vec{R}\times\vec{V}
&=&\,\mathrm{JacobiCN}\left[\frac{2}{R_0^2}(U+C),\frac{\sqrt{2}}{2}\right]\,.
\label{oscJvector}
\end{eqnarray}
whose square
fixes the constant in \eqref{Jsquare} to vanish. The {length} of \eqref{oscJvector} thus
oscillates as shown in FIG.\ref{Joscifig},
consistently with the breaking  of the axial symmetry.

\section{Approximate sandwich waves}\label{sandwichSec}
 
Realistic gravitational waves are (approximate) sandwich waves \cite{Hawking,BoPi89,
Zhang:2017rno,Zhang:2017geq,Steinbauer1997,Podolsky:1998in, PodSB, Zhang:2017jma,SBComment}, modelled by inserting the
profile into a Gaussian envelope \footnote{More general profiles could also be considered \cite{Hawking,Zhang:2017geq}.},
\begin{eqnarray}
ds_{Gauss}^2=\big(dX^2+dY^2+2dUdV\big)-
\frac{2}{\lambda}\exp\left[-\frac{U^2}{\lambda^2}\right]
\frac{\sin2\theta}{R^2}
\,dU^2\,.
\label{G-O21-GW}
\end{eqnarray}
The  parameter $\lambda$ rules the width of Gaussian bell. 
For  $\lambda \rightarrow \infty$ we recover the $U$-independent profile  \eqref{C10pot}, and 
 $\lambda \rightarrow 0$ is the impulsive limit it shrinks to $\delta(U)$  with sign alternating depending on the quadrant, though. The metric \eqref{G-O21-GW} is still a pp-wave however the $U$-dependent pre-factor breaks the $\Ort(2,1)$ symmetry.

\begin{figure}[h]
\includegraphics[scale=.35]{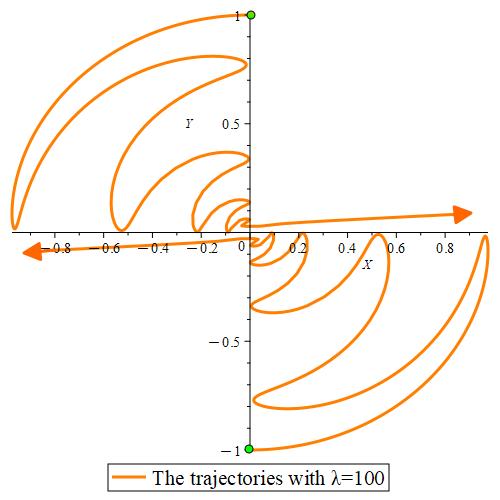}\qquad
\includegraphics[scale=.35]{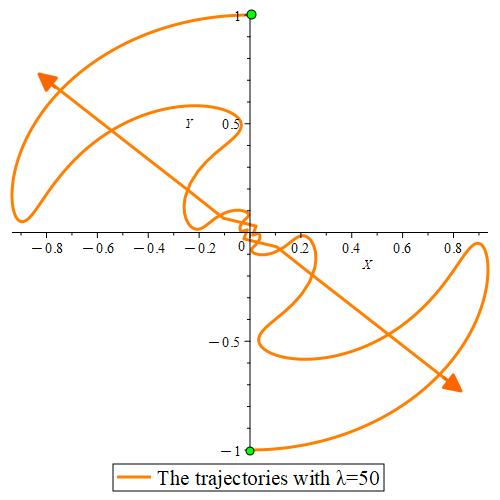}\\[10mm] 
\includegraphics[scale=.35]{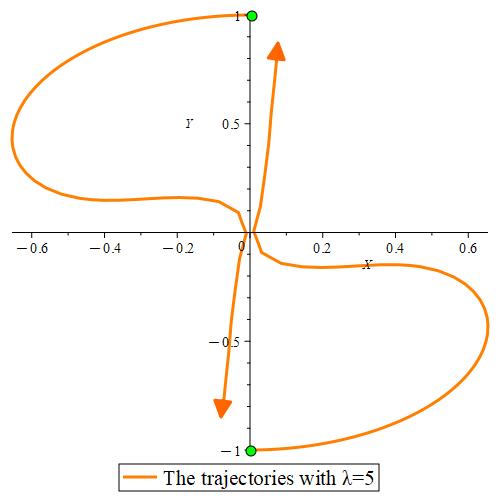}\qquad
\includegraphics[scale=.35]{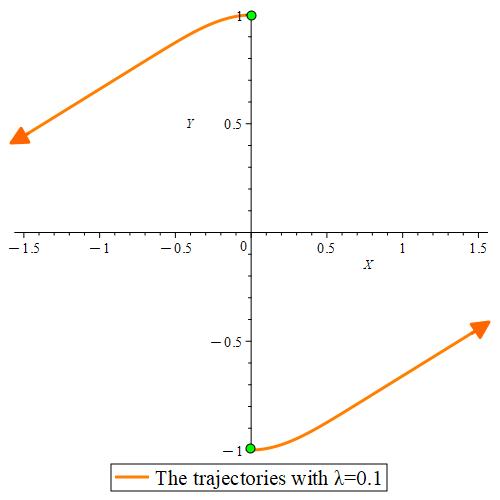}\\
\caption{\textit{\small The particle oscillates while shrinking towards the singularity. 
After getting close to the origin the particle, though, it turns suddenly back and gets \underline{expelled} along  an almost linear trajectory.} 
}
\label{shsandwich}
\end{figure}

We want to discover how do the periodic motions    \eqref{boundJacobi} behave. No analytic geodesics were found but or numerical calculations presented in FIG.\ref{shsandwich} show a peculiar behaviour.
All trajectories start form  \dgreen{$({\bf 0},\bf{\pm1})$} on the separating line between  the (originally)  \red{repulsive} or \cyan{attractive} quadrants. They follow initially  circular-looking trajectories, however after a while the damping due to the Gaussian starts to have its effect, and the particle oscillates closer and closer to the singularity. 
For large $\lambda$ the trajectory shrinks slowly, but with decreasing $\lambda$ the shrinking becomes more and more important. 

However the most dramatic effect is that
  after getting close to the origin the repulsive force wins, and the particle, instead of falling into the singularity, turns suddenly back and gets \emph{expelled} along an almost straight trajectory, consistently with the \emph{velocity effect} \cite{Grishchuk1989,Zhang:2018srn}.

\goodbreak

\bigskip

\gb{I wonder the reason of this turning back?}

\gb{can we say WHEN does it hapen ?}

\section{History: from Arnold through Newton, back to
Galilei}\label{historysec}

The zero - energy  motions  which oscillate  along quarter-of-circles in the attractive zone between the separation lines of the attractive and repulsive quarters have indeed quite remarkable ancestry. Let's proceed backwards in time.

We start with noting that our equations \eqref{C10eqs1} are reminiscent of the study of planetary motion by making use  of the Bohlin-Arnold duality between harmonic oscillators and the Kepler problem \cite{Bohlin,Arnold}.
 Eqns. \#(8.3) in \cite{ZHArnold} which \emph{assume} circular trajectories are consistent when the \emph{force is inversely proportional to the fifth power of the distance from the sun},
\beq
\text{force} \;\propto \; -\frac{1}{r^5}\,.
\label{inverse5}
\eeq
This facct was known already by Newton, who, in his  {\sl Principia}, inquired~: 
\emph{--~What force laws do allow for circular trajectories~?} -- and he found, using geometrical techniques that in addition to $r^{-2}$ one can have  also \eqref{inverse5}, 
 see \cite{Principia} vol. I Proposition VII. Problem II, where the proof is left as an exercise.  
 
Yet another intriguing feature is that both our circular solution in sec.\ref{PolarSec} and the parabolic trajectory 
 of the 1680 comet (discussed  by Newton in Book III Proposition XLI, Problem XXI of \cite{Principia}), has also \emph{zero energy}. These solutions separate bounded and unbounded motions.
 
Even more incredibly, FIG.4 in Galilei's {\sl Dialogo} \cite{Dialogo} written before Newton was even born, suggests circular motion which would pass through the center of the Earth.
 
Returning to  our circularly oscillating motions found in sect.\ref{PolarSec} we note that they  do \emph{not} enter into the Bohlin-Arnold framework. Let us explain. 
The Bohlin-Arnold trick \cite{Bohlin,Arnold} is based on a \emph{duality} between two central potentials proportional to $r^{a}$ and $r^{A}$, respectively, which are duals when the constraint
\beq
\left(1+\frac{a}{2}\right)\left(1+\frac{A}{2}\right)=1
\label{dualconstr}
\eeq
is satisfied; then motion in the $r^{a}$ and in the $r^{A}$ potentials can be swapped into each other. 

The newtonian potential corresponds, for example, to $a=-1$; its dual has therefore $A=2$ i.e., is an isotropic harmonic oscillator.  

The duality swaps also the dynamical symmetries of the oscillator with the Runge-Lenz vector-induced one of planetary motion. Working for simplicity in the plane using complex coordinates, $\zeta=\xi+i\eta$ for the oscillator and  $z=x+iy$  for the Kepler problem, the corresponding Levi-Civita - Bohlin - Arnold map \cite{LeviCivita,Bohlin,Arnold,Nersessian,ZHArnold},
\beq
z = \left(\zeta+\frac{1}{\zeta}\right)^2
\label{Arnoldmap}
\eeq
interchanges also those two sorts of dynamical symmetries \cite{Berard}.

The potential of the inverse-5 force \eqref{inverse5} is in turn self-dual, $a= A =-4$. 

However the \emph{inverse square} potential, which is precisely what we are interested in in this paper, has no Bohlin-Arnold dual: the constraint \eqref{dualconstr} can not be satisfied for $a=-2$. It is therefore a remarkable {\sl tour de force} that Sundaram et al \cite{Sundaram} could  extend the Bohlin-Arnold duality to that case. 

\section{Summary and discussions}\label{secV}

In this paper we study conformally related vacuum gravitational waves and their associated symmetries
by using a special M\"obius conformal transformation \eqref{ABCD-U}-\eqref{M-trans}. 
The vacuum condition is preserved by eliminating  the additional non-vacuum oscillator term \eqref{confprof} \cite{Masterov,ZZH2022}.
The resulting GW is in general different from the original one. The  transformation 
\eqref{ABCD-U}-\eqref{M-trans} carries a global GW into an (approximate) sandwich wave, as illustrated by LPP GW and CPP GW which exemplify also the memory effect \cite{ZelPol,Braginsky1985,Grishchuk1989,Ehlers,Sou73,Carroll4GW,Zhang:2017rno,Zhang:2017geq,EZHrev}.

A vacuum GW can also be invariant under the special M\"obius conformal transformation \eqref{ABCD-U}-\eqref{M-trans} when it has an $\Ort(2,1)$ symmetry. The remarkable efficiency of this symmetry comes from that its generators act on the radial variable only, therefore they apply equally well to anisotropic systems.

The particularly interesting example originating in molecular physics  \cite{Camblong:2001zt} but applied here in the gravitational context by using the Bargmann framework \cite{DBKP,Eisenhart,DGH91} is studied in some detail. It has the form of an anisotropic inverse-square potential \cite{AFF}. 

For the polar-molecular application, \eqref{polpot}, the familiar rotational symmetry is broken by an angle-dependent coefficient which makes it anisotropic: it alternates between repulsive and attractive  at every quarter-of-a circle, see \eqref{C10pot}. The particle is accordingly being pushed out to infinity or attracted towards the singularity at the origin, depending on the sign of the energy of the underlying non-relativistic problem. 
Bounded motion arise in the attractive quadrant, with  the particle oscillates along quarter-of-circle  between the lines which separate the attractive and repulsive quadrants.
Their behavior is reminiscent of that in the Kepler problem where the bounded (elliptical) and unbounded (hyperbolic) motions with  negative or positive energy are separated by zero-energy parabolic motions. 
 
\goodbreak

Analytic solutions were found also for escaping or incoming radial motion along the ``crests'' or ``valley bottoms" which correponds to the usual inverse-square potential with repulsive or attractive sign. 

The anisotropy breaks the rotational symmetry : the
 length of the angular momentum \eqref{oscJvector} 
 oscillates, as shown in FIG.\ref{Joscifig} in the periodic case.
  
The periodic motions in the attractive zone  
 show remarkable historical analogies, recounted in sec. \ref{historysec} by proceeding backwards in time.

The r\^ole played by the inverse-square potential in black-hole physics has been noticed before \cite{Claus} for the \emph{isotropic} Reissner-Nordstr\"om solution \cite{Camblong:2003mz}. 
The {anisotropic} metric \eqref{C10pot}, which seems to have escaped attention so far, is a pp wave which 
resembles that near  the ``Dirac String''  in the Lorentzian Taub-NUT metric \cite{Gibbons:2006gx,Holzegel:2006gn, GWG}.
\goodbreak


Replacing the trigonometric functions of $\theta$ in  \eqref{nosympot} or in \eqref{C10pot} by a constant, we would recover the familiar inverse-square profile 
\beq
ds^2 = \left(dR^2+R^2d\theta^2+ 2dUdV\right) - \frac{2}{R^2} dU^2,
\label{r-2metric}
\eeq
which is  reminiscent of Aichelburg-Sexl ultraboosts \cite{AichelburgSexl,Podolsky:1998rp,ASwiki},
\beq
ds^2 = \left(dr^2+r^2d\theta^2 + 2dudv\right) - 8\, \delta(u) {\log r} du^2,
\quad -\pi < \theta < \pi
\label{AiSeMetric}
\eeq
which describes the gravitational field of a massless particle which moves with the velocity of light. It can be considered as an approximation of the gravitational field of a photon \cite{ASwiki}. The metric \eqref{AiSeMetric} is indeed the impulsive limit of the axisymmetric Gaussian pulse
\beq
ds^2 = \left(dr^2+r^2d\theta^2+ 2dudv \right)- \frac{4a\, {\log r}}{\pi(1+a^2u^2)} du^2
\label{AiSeGauss}
\eeq
when $a \to \infty$.

The substantial difference between our inverse-square \eqref{r-2metric} and and the Aichelburg-Sexl metric \eqref{AiSeMetric} is
that the latter is a vacuum wave foutside the origin because of $\Delta(\log r) = \delta{(r)}$,
while \eqref{r-2metric} and our anisotropic generalisation \eqref{C10pot} are merely pp waves. 

The relation of the inverse-square metric with that of Aichelburg and Sexl can  be enlightend by putting \eqref{r-2metric} first into a Gaussian envelope, \eqref{G-O21-GW}, and then taking the impulsive limit $\lambda \to 0$. This yields 
an anisotropic analog of the Aichelburg-Sexl metric \eqref{AiSeMetric},
\beq
ds^2 = \left(dR^2+R^2d\theta^2+2dUdV\right) - \delta(U)\frac{2\sin 2\theta}{R^2} dU^2.
\label{impmol}
\eeq

Another difference is that our \eqref{nosympot} is $\orth(2,1)$-symmetric, while the Aichelburg-Sexl ultraboost is not~: $\log r$ is not scale-invariant which makes the discussion more elaborate.


\begin{acknowledgments}\vskip-4mm
We are grateful to Gary Gibbons for his insightful advices. Discussions are acknowledged also to Janos Balog and Mahmut Elbistan.
This work was partially supported by the National Natural Science Foundation of China (Grant No. 11975320).
\end{acknowledgments}


\end{document}